\documentclass[onecolumn, 11pt]{IEEEtran}
\PassOptionsToPackage{numbers, compress}{natbib}

\usepackage[utf8]{inputenc} 
\usepackage[T1]{fontenc}
\usepackage{url}
\usepackage{ifthen}
\usepackage{cite}
\usepackage[cmex10]{amsmath} % Use the [cmex10] option to ensure complicance
\usepackage{amsfonts}

\usepackage{algorithm,algorithmic}
\usepackage{color}
\usepackage{graphicx}

\DeclareRobustCommand{\bm}[1]{\mathbf{\boldsymbol{#1}}}
\DeclareRobustCommand{\deq}[0]{\overset{\rm d}{=}}

\begin{document}
\title{Macroscopic Analysis of Vector Approximate Message Passing in a Model Mismatch Setting}

%%% Single author, or several authors with same affiliation:
\author{%
  \IEEEauthorblockN{Takashi Takahashi and Yoshiyuki Kabashima\\}
  \IEEEauthorblockA{Department of Mathematical and Computing Science\\
                    Tokyo Institute of Technology\\
                    2-12-1, Ookayama, Meguro-ku, Tokyo, Japan\\
                    Email: takahashi.t.cc@m.titech.ac.jp, kaba@c.titech.ac.jp}
}

\maketitle

%%%%%%
%% Abstract: 
%% If your paper is eligible for the student paper award, please add
%% the comment "THIS PAPER IS ELIGIBLE FOR THE STUDENT PAPER
%% AWARD." as a first line in the abstract. 
%% For the final version of the accepted paper, please do not forget
%% to remove this comment!
%%
\begin{abstract}
    % THIS PAPER IS ELIGIBLE FOR THE STUDENT PAPER AWARD.
    Vector approximate message passing (VAMP) is an efficient approximate inference algorithm used for generalized linear models. Although VAMP exhibits excellent performance, particularly when measurement matrices are sampled from rotationally invariant ensembles, existing convergence and performance analyses have been limited mostly to cases in which the correct posterior distribution is available. Here, we extend the analyses for cases in which the correct posterior distribution is not used in the inference stage. We derive state evolution equations, which macroscopically describe the dynamics of VAMP, and show that their fixed point is consistent with the replica symmetric solution obtained by the replica method of statistical mechanics. We also show that the fixed point of VAMP can exhibit a microscopic instability, the critical condition of which agrees with that for breaking the replica symmetry. The results of numerical experiments support our findings.
    
\end{abstract}

%% The paper must be self-contained. However, if you are referring to
%% a full version for checking certain proofs, please provide the
%% publically accessible location below.  If the paper is completely
%% self-contained, you can remove the following line from your
%% submission.
% \textit{A full version of this paper is accessible at:}
% \url{http://isit2019.fr/} 

\section{Introduction} % introduction
\label{section:introduction}
% what's the problem?
Vector approximate message passing (VAMP) \cite{rangan2019vector, schniter2016vector, fletcher2018inference} is an approximate inference algorithm used for generalized linear models, which have many applications in signal processing, machine learning, and statistics. VAMP exhibits excellent convergence and inference performance, particularly when measurement matrices are sampled from rotationally invariant matrix ensembles. However, except for Gaussian channel setups, existing theoretical analyses have been limited to cases in which the correct posterior distribution is available. This type of consistency of the posterior is not necessarily satisfied in practical applications. Thus, we need to develop a methodology for analyzing VAMP used in a model mismatch setting, which is described in detail in the following section.

% our contribution
To answer this demand, we develop a method for analyzing the performance of VAMP in the model mismatch setting by utilizing techniques of statistical mechanics. We derive scalar state evolution (SE) equations, which describe the macroscopic dynamics of VAMP, and show that their fixed points are consistent with the replica symmetric (RS) solution obtained by the replica method of statistical mechanics. In addition, we show that the fixed point of VAMP can exhibit a microscopic instability, the critical condition of which accords with that for breaking the replica symmetry. 

% reminder of the paper
The remainder of this paper is organized as follows. In section \ref{section:setup}, we specify the problem setup treated in this study, and in section \ref{section:VAMP}, we briefly review VAMP. 
In section \ref{section:analysis}, we develop our methodology, and in section \ref{section:numerical experiment}, we demonstrate our findings through numerical experiments. We summarize and conclude our results in section \ref{section:summary and conclusion}.

%%%
\subsection{Additional related works}
\label{subsection:related works}
% Further related works for readers who are not familiar with AMP
VAMP belongs to the class of approximate message passing algorithms. This type of algorithm was first proposed as a computationally efficient code division multiple access multi-user detection algorithm \cite{kabashima2003cdma}. \cite{kabashima2003cdma} also revealed that its reconstruction performance is consistent with the optimal performance predicted by the replica method, which is believed to be exact in a large system limit. Subsequently, \cite{bolthausen2014iterative} developed its mathematically rigorous analysis. This rigorous analysis was further generalized in \cite{bayati2011dynamics, javanmard2013state}, which were extended to cases in which the correct posterior was not available and have been used in statistics and machine learning \cite{zdeborova2016statistical}. However, these optimality and rigorous guarantees require elements of measurement matrices to be i.i.d. zero-mean Gaussians, which are not realistic in practical situations. VAMP and similar generalizations have extended the guarantees from i.i.d. Gaussian measurement matrices to the larger class of rotationally invariant random matrices \cite{rangan2019vector, schniter2016vector, fletcher2018inference, ma2017orthogonal, ma2014turbo}. However, existing analyses of them have been conducted mostly under the assumption that the correct posterior is available.

%%%%%
\section{Setup}% problem setup treated in this paper
\label{section:setup}
% setup (signal estimation, actual data generation process)
We address the problem of estimating a signal vector $\bm{x}_0 = (x_1,\dots,x_N)^\top \in \mathbb{R}^N$ from observation data $\bm{y} = (y_1, \dots, y_M)^\top \in \mathbb{R}^M$, which is generated from the following \emph{actual} measurement process:
\begin{align}
    \bm{y} \sim q_{\bm{y}| \bm{z}} &= \prod_{\mu=1}^M q_{y | z}, 
    % \label{eq:actual channel}
    \nonumber 
    % \\
    \quad
    \bm{z} = A\bm{x}_0, \\
    \bm{x}_0 &\sim q_{\bm{x}_0}=\prod_{i=1}^N q_{x_{0}},
    % \label{eq:actual prior}
    \nonumber
\end{align}
where $\top$ denotes vector/matrix transpose, $A=[a_{\mu i}]\in\mathbb{R}^{M\times N}$ is a known measurement matrix, $q_{\bm{y}| \bm{z}}$ is an actual measurement channel, and $q_{\bm{x}_0}$ is an actual prior. In this study, we assume that $A$ is drawn from rotationally invariant random matrix ensembles. Specifically, for the singular value decomposition of $A=USV^\top, U\in\mathbb{R}^{M\times M}, S\in\mathbb{R}^{M\times N}, V\in\mathbb{R}^{N\times N}$, we assume $U$ and $V$ are drawn from uniform distributions over $M\times M$ and $N\times N$ orthogonal matrices.

% postulated data generation process and estimation procedure
In the estimation stage, we consider generalized Bayesian inference using a \emph{mismatched} model $p_{\bm{y}|\bm{z}}=\prod_{\mu=1}^M p_{y|z} \ne q_{\bm{y}|\bm{x}}$ and/or $p_{\bm{x}_0}=\prod_{i=1}^N p_{x_0} \ne q_{\bm{x}_0}$ as: 
\begin{align}
    % \vspace{-10pt}
    \hat{\bm{x}} &= \int \bm{x} p(\bm{x}, \bm{z})d\bm{x}d\bm{z}, \label{eq:estimator} \\
    p(\bm{x}, \bm{z}) &= \frac{1}{Z}p_{\bm{y}|\bm{z}}(\bm{y}| \bm{z})^\beta p_{\bm{x}_0}(\bm{x})^\beta \delta(A\bm{x} - \bm{z}), 
    \label{eq:boltzmann distribution}
    \\
    Z &= \int p_{\bm{y}|\bm{z}}(\bm{y}| \bm{z})^\beta p_{\bm{x}_0}(\bm{x})^\beta \delta(A\bm{x} - \bm{z}) d\bm{x}d\bm{z}.
    \label{eq:partition function}
\end{align}
\vspace{-10pt}

\noindent Here, $\hat{\bm{x}}$ is the estimator of $\bm{x}_0$. In addition, the positive parameter $\beta$ is introduced for handling minimum mean squared error (MMSE) ($\beta=1$) and maximum a posteriori ($\beta\to \infty$) estimators in a unified manner. The normalization factor $Z$ plays a critical role in the analysis of statistical mechanics. 

%%%%%
\section{Vector approximate message passing}  % review of the vector approximate message passing
\label{section:VAMP}

%%%%% Algorithm: VAMP %%%%%
\begin{algorithm}[t]
\caption{VAMP}
\begin{algorithmic}[1]  \label{algo:VAMP}
\REQUIRE{
    Denoisers $\bm{g}_{1x}, \bm{g}_{1z}$ from (\ref{eq:denoiser}),  LMMSE estimators $\bm{g}_{2x}, \bm{g}_{2z}$ from (\ref{eq:LMMSE estimator x}) and (\ref{eq:LMMSE estimator z}), and number of iterations $T_{\rm iter}$.
}
\STATE{Select initial $\bm{h}_{1x}^{(1)}, \bm{h}_{1z}^{(1)}, \hat{Q}_{1x}^{(1)}>0, \hat{Q}_{1z}^{(1)}>0.$}
\FOR{for $t=1,2,\dots, T_{\rm iter}$}
    \STATE{// Factorized part}
    \vspace{1.5pt}
    \STATE{$\hat{\bm{x}}_1^{(t)} = \bm{g}_{1x}(\bm{h}_{1x}^{(t)}, \hat{Q}_{1x}^{(t)}), \chi_{1x}^{(t)} = \langle \bm{g}_{1x}^\prime(\bm{h}_{1x}^{(t)}, \hat{Q}_{1x}^{(t)}) \rangle.$
    \label{line:factorized x}
    }
    \vspace{1.5pt}
    \STATE{$\hat{\bm{z}}_1^{(t)} = \bm{g}_{1z}(\bm{h}_{1z}^{(t)}, \hat{Q}_{1z}^{(t)}), \chi_{1z}^{(t)} = \langle \bm{g}_{1z}^\prime(\bm{h}_{1z}^{(t)}, \hat{Q}_{1z}^{(t)}) \rangle.$
    \label{line:factorized z}}
    \vspace{1.5pt}
    \STATE{// message passing}
    \vspace{1.5pt}
    \STATE{$\bm{h}_{2x}^{(t)} = \hat{\bm{x}}_1^{(t)}/\chi_{1x}^{(t)} - \bm{h}_{1x}^{(t)},\;\bm{h}_{2z}^{(t)} = \hat{\bm{z}}_1^{(t)}/\chi_{1z}^{(t)} - \bm{h}_{1z}^{(t)}.$
    \label{line:message FtoG 1}
    }
    \vspace{1.5pt}
    \STATE{$\hat{Q}_{2x}^{(t)} = 1/\chi_{1x}^{(t)} - \hat{Q}_{1x}^{(t)}, \; \hat{Q}_{2z}^{(t)}=1/\chi_{1z}^{(t)} - \hat{Q}_{1z}^{(t)}.$
    \label{line:message FtoG 2}}
    \vspace{1.5pt}
    \STATE{// Gaussian part}
    \vspace{1.5pt}
    \STATE{$\hat{\bm{x}}_2^{(t)} = \bm{g}_{2x}(\bm{h}_{2x}^{(t)}, \bm{h}_{2z}^{(t)}, \hat{Q}_{2x}^{(t)}, \hat{Q}_{2z}^{(t)}).$
    \label{line:Gaussian x 1}
    }
    \vspace{1.5pt}
    \STATE{$\chi_{2x}^{(t)} = \langle \bm{g}_{2x}^\prime(\bm{h}_{2x}^{(t)}, \bm{h}_{2z}^{(t)}, \hat{Q}_{2x}^{(t)}, \hat{Q}_{2z}^{(t)}) \rangle.$
    \label{line:Gaussian x 2}
    }
    \vspace{1.5pt}
    \STATE{$\hat{\bm{z}}_2^{(t)} = \bm{g}_{2z}(\bm{h}_{2x}^{(t)}, \bm{h}_{2z}^{(t)}, \hat{Q}_{2x}^{(t)}, \hat{Q}_{2z}^{(t)}).$
    \label{line:Gaussian z 1}
    }
    \vspace{1.5pt}
    \STATE{$\chi_{2z}^{(t)} = \langle \bm{g}_{2z}^\prime(\bm{h}_{2x}^{(t)}, \bm{h}_{2z}^{(t)}, \hat{Q}_{2x}^{(t)}, \hat{Q}_{2z}^{(t)}) \rangle.$
    \label{line:Gaussian z 2}}
    \vspace{1.5pt}
    \STATE{// message passing}
    \vspace{1.5pt}
    \STATE{$\bm{h}_{1x}^{(t+1)} = \hat{\bm{x}}_2^{(t)}/\chi_{2x}^{(t)} - \bm{h}_{2x}^{(t)},\; \bm{h}_{1z}^{(t+1)} = \hat{\bm{z}}_2^{(t)}/\chi_{2z}^{(t)} - \bm{h}_{2z}^{(t)}.$
    \label{line:message GtoF 1}}
    \vspace{1.5pt}
    \STATE{$\hat{Q}_{1x}^{(t+1)} = 1/\chi_{2x}^{(t)} - \hat{Q}_{2x}^{(t)},\; \hat{Q}_{1z}^{(t+1)}=1/\chi_{2z}^{(t)} - \hat{Q}_{2z}^{(t)}.$
    \label{line:message GtoF 2}}
    % \vspace{1.5pt}
    % \STATE{$.$}
    % \vspace{1.5pt}
    % \STATE{$\hat{Q}_{1z}^{(t+1)}=1/\chi_{2z}^{(t)} - \hat{Q}_{2z}^{(t)}.$}
\ENDFOR
\RETURN{$\hat{\bm{x}}_1^{(T_{\rm iter})}$.}

\end{algorithmic}
\end{algorithm}
%%%%%%%%%%%%%%%%%%%%%%%%%%%%%%%%%%%

% definition of VAMP (already derived by Rangan et al. 2016)
Algorithm \ref{algo:VAMP} shows VAMP to evaluate the estimator (\ref{eq:estimator}) \cite{schniter2016vector}.
There, $\bm{g}_{1x}, \bm{g}_{1z}$ are termed \emph{denoising functions} defined as: 
\begin{align}
    \bm{g}_{1x}(\bm{h}_{x},\hat{Q}_{x}) = \frac{\partial \tilde{\phi}_{x}(\bm{h}_{x}, \hat{Q}_{x})}{\partial \bm{h}_x},\, 
    \bm{g}_{1z}(\bm{h}_{z},\hat{Q}_{z}) = \frac{\partial \tilde{\phi_z}(\bm{h}_z, \hat{Q}_z)}{\partial \bm{h}_z} , 
    \label{eq:denoiser}
\end{align}
where 
\begin{align*}
    \tilde{\phi}_{x}(\bm{h}_{x},\hat{Q}_{x}) &= \frac{1}{\beta}\log\int \prod_{i=1}^N e^{-\frac{\beta\hat{Q}_{x}}{2}x_i^2 + \beta h_{x,i} x_i}p_{x_0}(x_i)^{\beta}d\bm{x}, 
    \\
    \tilde{\phi}_z (\bm{h}_{z},\hat{Q}_{z}) &= \frac{1}{\beta}\log\int\prod_{\mu=1}^M  e^{-\frac{\beta \hat{Q}_z}{2}z_\mu^2 + \beta h_{z,\mu}z_\mu}p_{y|z}(y_\mu | z_\mu)^\beta d\bm{z}.
\end{align*}
In addition, $\bm{g}_{2x}, \bm{g}_{2z}$ are termed linear MMSE (LMMSE) estimators defined as:
\begin{align}
    \bm{g}_{2x}(\bm{h}_{x}, \bm{h}_{z}, \hat{Q}_{x},\hat{Q}_{z}) &= K^{-1}(\bm{h}_{x} + A^\top \bm{h}_z), 
    \label{eq:LMMSE estimator x}
    \\
    \bm{g}_{2z}(\bm{h}_{x}, \bm{h}_{z}, \hat{Q}_{x},\hat{Q}_{z}) &= A\bm{g}_{2x}(\bm{h}_{x}, \bm{h}_{z}, \hat{Q}_{x},\hat{Q}_{z}), 
    \label{eq:LMMSE estimator z}
\end{align}
where $K= \hat{Q}_xI_N + \hat{Q}_zA^\top A$. In addition, $\langle \bm{g}^\prime_{kx}(\bm{h}_{kx}, \hat{Q}_{kx})\rangle$, $\langle \bm{g}^\prime_{kz}(\bm{h}_{kz}, \hat{Q}_{kz})\rangle, k=1,2$ are defined as: 
\begin{align}
    \langle \bm{g}^\prime_{kx}(\bm{h}_{kx}, \hat{Q}_{kx})\rangle &= \frac{1}{N}{\rm Tr}\frac{\partial \bm{g}_{kx}(\bm{h}_{kx}, \hat{Q}_{kx})}{\partial \bm{h}_{kx}}, 
    \nonumber 
    \\
    \langle \bm{g}^\prime_{kz}(\bm{h}_{kz}, \hat{Q}_{kz})\rangle &= \frac{1}{M}{\rm Tr}\frac{\partial \bm{g}_{kz}(\bm{h}_{kz}, \hat{Q}_{kz})}{\partial \bm{h}_{kz}}.
    \nonumber
\end{align}
At a fixed point, $\hat{\bm{x}}_1^{(t)}=\hat{\bm{x}}_2^{(t)}$, $\hat{\bm{z}}_1^{(t)}=\hat{\bm{z}}_2^{(t)}$, $\chi_{1x}^{(t)}=\chi_{2x}^{(t)}$, and $\chi_{1z}^{(t)}=\chi_{2z}^{(t)}$ are achieved by construction.

%%%%%
\section{Analysis}  % out analysis
\label{section:analysis}

%%%
\subsection{SE}
% assumptions to derive SE
In this section, we derive SE of VAMP in the limit $N,M\to\infty, M/N=\delta\in(0,\infty)$.
For this, we make the following assumption:
\paragraph*{Assumption}
At each iteration $t=1,2,\dots ,T_{\rm iter}$, positive constants $\hat{m}_{kx}^{(t)}$, $\hat{m}_{kz}^{(t)}$, $\hat{\chi}_{kx}^{(t)}$, $\hat{\chi}_{kz}^{(t)}\in\mathbb{R}$, $(k=1,2)$ exist such that for the singular value decomposition $A=USV^\top$, 
\begin{align}
    \bm{h}_{1x}^{(t)} - \hat{m}_{1x}^{(t)}\bm{x}_0 &\deq \sqrt{\hat{\chi}_{1x}^{(t)}}\bm{\xi}_{1x}^{(t)},
    \nonumber
    \\
    \bm{h}_{1z}^{(t)} - \hat{m}_{1z}^{(t)}\bm{z}_0 &\deq \sqrt{\hat{\chi}_{1z}^{(t)}}\bm{\xi}_{1z}^{(t)}, 
    \nonumber
    \\
    V^\top (\bm{h}_{2x}^{(t)} - \hat{m}_{2x}^{(t)}\bm{x}_0) &\deq \sqrt{\hat{\chi}_{2x}^{(t)}}\bm{\xi}_{2x}^{(t)}, 
    \nonumber
    \\
    U^\top (\bm{h}_{2z}^{(t)} - \hat{m}_{2z}^{(t)}\bm{z}_0) &\deq \sqrt{\hat{\chi}_{2z}^{(t)}}\bm{\xi}_{2z}^{(t)},
    \nonumber
\end{align}
hold, where $\deq$ denotes equality of empirical distributions $\bm{z}_0=A\bm{x}_0$, and $\bm{\xi}_{kx}^{(t)}, \bm{\xi}_{kz}^{(t)}, (k=1,2, t=1,2,\dots ,T_{\rm iter})$ are mutually independent standard Gaussian variables. We also assume that these are independent from $\bm{x}_0, \bm{z}_0, V^\top \bm{x}_0$, and $U^\top \bm{z}_0$.
Although the empirical distributions of $\bm{h}_{2x}^{(t)}-\hat{m}_{2x}^{(t)}\bm{x}_0$ and $\bm{h}_{2z}^{(t)}-\hat{m}_{2z}^{(t)}\bm{z}_0$ are generally different from Gaussian, the rotational invariance of $A$ makes this assumption plausible. 

% derivation of SE
To characterize the macroscopic behavior of VAMP, we introduce the following macroscopic variables: $m_{kx}^{(t)} = \bm{x}_0^\top \hat{\bm{x}}_k^{(t)}/N$, $q_{kx}^{(t)} = \|\hat{\bm{x}}_{k}^{(t)}\|_2^2/N$, $m_{kz}^{(t)} = \bm{z}_0^\top \hat{\bm{z}}_k^{(t)}/M$, $q_{kz} = \|\hat{\bm{z}}_{k}^{(t)}\|_2^2/M$, $k=1,2$, $T_x = \|\bm{x}_0\|_2^2/N$, $T_z = \|\bm{z}_0\|_2^2/M$.  Then, under the aforementioned assumption, VAMP indicates that $\hat{m}_{2x}^{(t)}$, $\hat{\chi}_{2x}^{(t)}$ are written by $\hat{m}_{1x}^{(t)}$, $\hat{\chi}_{1x}^{(t)}$, $\chi_{1x}^{(t)}$, $m_{1x}^{(t)}$, $q_{1x}^{(t)}$ and $T_x$ as follows:
\begin{align}
    \hat{m}_{2x}^{(t)} &= \frac{\bm{x}_0^\top \bm{h}_{2x}^{(t)}}{\|\bm{x}_0\|_2^2} = \frac{\bm{x}_0^\top }{\|\bm{x}_0\|_2^2}
    \left(
        \frac{\hat{\bm{x}}_1^{(t)}}{\chi_{1x}^{(t)}}-\bm{h}_{1x}^{(t)}
    \right)
    % \nonumber 
    % \\
    % &=
    =
    \frac{m_{1x}^{(t)}}{T_x\chi_{1x}^{(t)}} - \hat{m}_{1x}^{(t)},
    \nonumber 
% \end{align}
% and 
% \begin{align}
    \nonumber \\
    \hat{\chi}_{2x}^{(t)} &= \frac{1}{N}\|\bm{h}_{2x}^{(t)} - \hat{m}_{2x}^{(t)}\bm{x}_0\|_2^2,
    \nonumber 
    \\
    &=
    \frac{1}{N}
    \left\| 
        \frac{\hat{\bm{x}}_1^{(t)}}{\chi_{1x}^{(t)}} - \frac{m_{1x}^{(t)}}{T_x\chi_{1x}^{(t)}}\bm{x}_0 - (\bm{h}_{1x}^{(t)} - \hat{m}_{1x}^{(t)}\bm{x}_0)
    \right\|_2^2
    \nonumber 
    \\
    &= \frac{q_{1x}^{(t)}}{(\chi_{1x}^{(t)})^2} - \frac{(m_{1x}^{(t)})^2}{T_x(\chi_{1x}^{(t)})^2} + \hat{\chi}_{1x}^{(t)} - 2\frac{(\bm{h}_{1x}^{(t)}- \hat{m}_{1x}^{(t)}\bm{x}_0)^\top \hat{\bm{x}}_1^{(t)}}{N\chi_{1x}^{(t)}}
    + 2
        \frac{m_{1x}^{(t)}}{T_x \chi_{1x}^{(t)}} \frac{(\bm{h}_{1x}^{(t)}- \hat{m}_{1x}^{(t)}\bm{x}_0)^\top \bm{x}_0}{N}
    \label{eq:chi_2x_hat}
    \\
    & \xrightarrow{N\to\infty} \frac{q_{1x}^{(t)}}{(\chi_{1x}^{(t)})^2} - \frac{(m_{1x}^{(t)})^2}{T_x(\chi_{1x}^{(t)})^2} - \hat{\chi}_{1x}^{(t)}
    .
    \nonumber 
\end{align}
In (\ref{eq:chi_2x_hat}), we replaced the last two terms with  $-2\hat{\chi}_{1x}^{(t)}$ and $0$ using the assumption and an identity 
$\int xf(x)Dx = \int f^\prime (x) Dx$
for $\forall{f}(x)$, where $Dx=e^{-x^2/2}/\sqrt{2\pi} dx$. 
Similarly,  $\hat{m}_{1x}^{(t+1)}$ and $\hat{\chi}_{1x}^{(t+1)}$ are written by $\hat{m}_{2x}^{(t)}$, $\hat{\chi}_{2x}^{(t)}$, $\chi_{2x}^{(t)}$, $m_{2x}^{(t)}$, $q_{2x}^{(t)}$, and $T_x$ as follows:

\vspace{-15pt}
\begin{align}
    \hat{m}_{1x}^{(t+1)} &=
    \frac{m_{2x}^{(t)}}{T_x\chi_{2x}^{(t)}} - \hat{m}_{2x}^{(t)},
    \nonumber 
    \\
    \hat{\chi}_{1x}^{(t+1)} 
    &= \frac{q_{2x}^{(t)}}{(\chi_{2x}^{(t)})^2} - \frac{(m_{2x}^{(t)})^2}{T_x (\chi_{2x}^{(t)})^2} 
    % \nonumber 
    % \\
    % &
    -\frac{2}{\chi_{2x}^{(t)}}\frac{ \left\{V^\top(\bm{h}_{2x} - \hat{m}_{2x}^{(t)}\bm{x}_0)\right\}^\top \left(V^\top \hat{\bm{x}}_2^{(t)}\right)}{N}
    % \nonumber \\
    % &
    +\frac{2m_{2x}^{(t)}}{T_x \chi_{2x}^{(t)}}\frac{ \left\{V^\top(\bm{h}_{2x} - \hat{m}_{2x}^{(t)}\bm{x}_0)\right\}^\top \left(V^\top \hat{\bm{x}}_0\right)}{N}
    \nonumber 
    \\
    &\xrightarrow{N\to\infty} \frac{q_{2x}^{(t)}}{(\chi_{2x}^{(t)})^2} - \frac{(m_{2x}^{(t)})^2}{T_x(\chi_{2x}^{(t)})^2} - \hat{\chi}_{2x}^{(t)}
    .
    \nonumber 
\end{align}
% \vspace{-10pt}

% state evolution
\noindent A similar argument applies to $\hat{m}_{1z}^{(t)}, \hat{\chi}_{1z}^{(t)}, \hat{m}_{2z}^{(t)}$, and $\hat{\chi}_{2z}^{(t)}$. Thus, the SE of Algorithm \ref{algo:VAMP} is expressed as follows:

\noindent\emph{Initialization:} initialize $\hat{Q}_{1x}^{(1)}, \hat{Q}_{1z}^{(1)}, \hat{m}_{1x}^{(1)}, \hat{m}_{1z}^{(1)}, \hat{\chi}_{1x}^{(1)}, \hat{\chi}_{1z}^{(1)}>0$.

\noindent\emph{Factorized part:}
\begin{align}
    %% F estimation
    % x
    m_{1x}^{(t)} &= \int \frac{\partial \phi_x}{\partial \hat{m}_{1x}^{(t)}}q_{x_0}(x_0)dx_0 D\xi_{1x},
    \label{eq:factorized x 1}
    \\
    \chi_{1x}^{(t)} &= \int \frac{\partial^2 \phi_x}{\partial (\sqrt{\hat{\chi}_{1x}^{(t)}}\xi_{1x})^2}q_{x_0}(x_0)dx_0 D\xi_{1x},
    \label{eq:factorized x 2}
    \\
    q_{1x}^{(t)} &= \int 
        \left[
            \frac{\partial \phi_x}{\partial (\sqrt{\hat{\chi}_{1x}^{(t)}}\xi_{1x})}
        \right]^2q_{x_0}(x_0)dx_0 D\xi_{1x},
    % \label{eq:factorized x 3}
    \label{eq:factorized x 3}
    \\
    % z
    m_{1z}^{(t)} &= \int \frac{\partial \phi_z}{\partial \hat{m}_{1z}^{(t)}}\sqrt{\frac{\hat{T}_z}{2\pi}}e^{-\frac{\hat{T}_z}{2}z_0^2}q_{y|z}(y|z_0)dy dz_0 D\xi_{1z},
    \label{eq:factorized z 1}
    \\
    \chi_{1z}^{(t)} &= \int \frac{\partial^2 \phi_z}{\partial (\sqrt{\hat{\chi}_{1z}^{(t)}}\xi_{1z})^2}\sqrt{\frac{\hat{T}_z}{2\pi}}e^{-\frac{\hat{T}_z}{2}z_0^2}q_{y|z}(y|z_0)dy dz_0 D\xi_{1z},
    \label{eq:factorized z 2}
    \\
    q_{1z}^{(t)} &= \int \left[
            \frac{\partial \phi_z}{\partial (\sqrt{\hat{\chi}_{1x}^{(t)}}\xi_{1z})}
        \right]^2
    \sqrt{\frac{\hat{T}_z}{2\pi}}e^{-\frac{\hat{T}_z}{2}z_0^2}q_{y|z}(y|z_0)dy dz_0 D\xi_{1z},
    \label{eq:factorized z 3}
\end{align}
\emph{Message passing:}
\begin{align}
    %
    %% F to G
    % x
    \hat{Q}_{2x}^{(t)} &= \frac{1}{\chi_{1x}^{(t)}} - \hat{Q}_{1x}^{(t)} ,\quad \hat{Q}_{2z}^{(t)} = \frac{1}{\chi_{1z}^{(t)}} - \hat{Q}_{1z}^{(t)}
    \label{eq:message FtoG 1}
    \\
    \hat{m}_{2x}^{(t)} &= \frac{m_{1x}^{(t)}}{T_x \chi_{1x}^{(t)}} - \hat{m}_{1x}^{(t)} , \quad \hat{m}_{2z}^{(t)} = \frac{m_{1z}^{(t)}}{T_z \chi_{1z}^{(t)}} - \hat{m}_{1z}^{(t)} ,
    \label{eq:message FtoG 2}
    \\
    \hat{\chi}_{2x}^{(t)} &= \frac{q_{1x}^{(t)}}{(\chi_{1x}^{(t)})^2} - \frac{(\hat{m}_{1x}^{(t)})^2}{T_x (\chi_{1x}^{(t)})^2} -  \hat{\chi}_{1x}^{(t)},
    \label{eq:message FtoG 3}
    \\
    \hat{\chi}_{2z}^{(t)} &= \frac{q_{1z}^{(t)}}{(\chi_{1z}^{(t)})^2} - \frac{(\hat{m}_{1z}^{(t)})^2}{T_z (\chi_{1z}^{(t)})^2} -  \hat{\chi}_{1z}^{(t)},
    \label{eq:message FtoG 4}
\end{align}
\emph{Gaussian part:}
\begin{align}
    % x
    m_{2x}^{(t)} &= T_x \mathbb{E}_\lambda
    \left[
        \frac{
            \hat{m}_{2x}^{(t)} + \lambda \hat{m}_{2z}^{(t)}
        }{
            \hat{Q}_{2x}^{(t)} + \lambda \hat{Q}_{2z}^{(t)}
        }
    \right],
    \label{eq:gaussian x 1}
    \\
    \chi_{2x}^{(t)} &= \mathbb{E}_\lambda\left[\frac{1}{\hat{Q}_{2x}^{(t)} + \lambda \hat{Q}_{2z}^{(t)}}\right],
    \label{eq:gaussian x 2}
    \\
    q_{2x}^{(t)} &\hspace{-2pt}=\hspace{-2pt} \mathbb{E}_\lambda\hspace{-2pt}
    \left[\hspace{-2pt}
        \frac{\hat{\chi}_{2x}^{(t)} + \lambda\hat{\chi}_{2z}^{(t)}}{
            (
                \hat{Q}_{2x}^{(t)} + \lambda\hat{Q}_{2z}^{(t)}
            )^2}
    \hspace{-3pt}\right]
    \hspace{-2pt}+\hspace{-2pt}T_x
    \mathbb{E}_\lambda\hspace{-2pt}
    \left[\hspace{-1pt}
        \frac{
            (
                \hat{m}_{2x}^{(t)} + \lambda\hat{m}_{2z}^{(t)}
            )^2
        }{
            (
                \hat{Q}_{2x}^{(t)} + \lambda\hat{Q}_{2z}^{(t)}
            )^2
        }
    \hspace{-2pt}\right]\hspace{-3pt},
    \label{eq:gaussian x 3}
    \\
    % z
    m_{2z}^{(t)} &= \frac{T_x}{\delta} \mathbb{E}_\lambda
    \left[
        \frac{
            \lambda(\hat{m}_{2x}^{(t)} + \lambda \hat{m}_{2z}^{(t)})
        }{
            \hat{Q}_{2x}^{(t)} + \lambda \hat{Q}_{2z}^{(t)}
        }
    \right],
    \label{eq:gaussian z 1}
    \\
    \chi_{2z}^{(t)} &= \frac{1}{\delta}\mathbb{E}_\lambda\left[\frac{\lambda}{\hat{Q}_{2x}^{(t)} + \lambda \hat{Q}_{2z}^{(t)}}\right],
    \label{eq:gaussian z 2}
    \\
    q_{2z}^{(t)} &= \frac{1}{\delta}\mathbb{E}_\lambda\hspace{-2pt}
    \left[\hspace{-1pt}
        \frac{\lambda(\hat{\chi}_{2x}^{(t)} + \lambda\hat{\chi}_{2z}^{(t)})}{
            (
                \hat{Q}_{2x}^{(t)} + \lambda\hat{Q}_{2z}^{(t)}
            )^2}
    \hspace{-1pt}\right]
    \hspace{-2pt}+ \hspace{-2pt}\frac{T_x}{\delta}
    \mathbb{E}_\lambda\hspace{-2pt}
    \left[\hspace{-1pt}
        \frac{
            \lambda(
                \hat{m}_{2x}^{(t)} + \lambda\hat{m}_{2z}^{(t)}
            )^2
        }{
            (
                \hat{Q}_{2x}^{(t)} + \lambda\hat{Q}_{2z}^{(t)}
            )^2
        }
    \hspace{-1pt}\right],
    \label{eq:gaussian z 3}
\end{align}
\emph{Message Passing:}
\begin{align}
    \hat{Q}_{1x}^{(t+1)} &= \frac{1}{\chi_{2x}^{(t)}} - \hat{Q}_{2x}^{(t)} ,
    \quad
    \hat{Q}_{1z}^{(t+1)} = \frac{1}{\chi_{2z}^{(t)}} - \hat{Q}_{2z}^{(t)}
    \label{eq:message GtoF 1}
    \\
    \hat{m}_{1x}^{(t+1)} &= \frac{m_{2x}^{(t)}}{T_x \chi_{2x}^{(t)}} - \hat{m}_{2x}^{(t)} , 
    \quad
    \hat{m}_{1z}^{(t+1)} = \frac{m_{2z}^{(t)}}{T_z \chi_{2z}^{(t)}} - \hat{m}_{2z}^{(t)} ,
    \label{eq:message GtoF 2}
    \\
    \hat{\chi}_{1x}^{(t+1)} &= \frac{q_{2x}^{(t)}}{(\chi_{2x}^{(t)})^2} - \frac{(\hat{m}_{2x}^{(t)})^2}{T_x (\chi_{2x}^{(t)})^2} -  \hat{\chi}_{2x}^{(t)},
    \label{eq:message GtoF 3}
    \\
    \hat{\chi}_{1z}^{(t+1)} &= \frac{q_{2z}^{(t)}}{(\chi_{2z}^{(t)})^2} - \frac{(\hat{m}_{2z}^{(t)})^2}{T_z (\chi_{2z}^{(t)})^2} -  \hat{\chi}_{2z}^{(t)},
    % \label{eq:message GtoF 4}
    \label{eq:message GtoF 4}
\end{align}
where $t=1,2,\dots,T_{\rm iter}$, $\mathbb{E}_\lambda[\dots ]$ is an average over the limiting eigenvalue spectrum of $A^\top A$, 
\begin{align}
    \phi_x(\hat{m}_x, \hat{Q}_x, \hat{\chi}_x;x_0)= \frac{1}{\beta}\log\int e^{-\frac{\beta\hat{Q}_x}{2}x^2 + \beta(\hat{m}_x x_0 + \sqrt{\hat{\chi}_x}\xi_x)x}p_{x_0}(x)dx,
    \nonumber \\
    \phi_z(\hat{m}_z, \hat{Q}_z, \hat{\chi}_z;z_0) = \frac{1}{\beta} \log \int e^{-\frac{\beta\hat{Q}_z}{2}z^2 + \beta(\hat{m}_z z_0 + \sqrt{\hat{\chi}_z}\xi_z)z}p_{y|z}(y| z)dz,
    \nonumber
\end{align}
$T_x=\int x_0^2 q_{x_0}(x_0)dx_0$, $T_z= \delta^{-1}\mathbb{E}_\lambda[\lambda]T_x$, and $\hat{T}_z = 1/T_z$.
In (\ref{eq:factorized x 1})-(\ref{eq:factorized z 3}), the functions $\phi_x$ and $\phi_z$ are evaluated at $(\hat{m}_{1x}^{(t)}, \hat{Q}_{1x}^{(t)}, \hat{\chi}_{1x}^{(t)}; x_0)$, and $(\hat{m}_{1z}^{(t)}, \hat{Q}_{1z}^{(t)}, \hat{\chi}_{1z}^{(t)}; z_0)$.
At the fixed point, $\chi_{1x}^{(t)} = \chi_{2x}^{(t)} = \chi_x$, $q_{1x}^{(t)} = q_{2x}^{(t)} =q_x$, $m_{1x}^{(t)} = m_{2x}^{(t)} = m_x$, $\chi_{1z}^{(t)} = \chi_{2z}^{(t)} = \chi_z$, $q_{1z}^{(t)} = q_{2z}^{(t)} =q_z$, and $m_{1z}^{(t)} = m_{2z}^{(t)} = m_z$ are achieved. This is the first result of this study.

% comment on the derived SE equations
Two points are noteworthy here. First, the aforementioned SE cannot be written using only the MSE $\|\bm{x}_0-\hat{\bm{x}}_1^{(t)}\|_2^2/N = T_x - 2m_{1x}^{(t)} + q_{1x}^{(t)}$. This point is strikingly different from the existing SE of VAMP \cite{rangan2019vector}. The second point concerns the meaning of the macroscopic variables. At the fixed point, $m_x, \chi_x, q_x$ are approximate values of $\bm{x}_0^\top \hat{\bm{x}}/N$, $\beta(\int \|\bm{x}\|_2^2p(\bm{x},\bm{z})d\bm{x}d\bm{z} - \|\hat{\bm{x}}\|_2^2)/N$, and $\|\hat{\bm{x}}\|_2^2/N$, respectively. Similarly, let $\bm{\hat{z}}$ be $\int \bm{z}p(\bm{x},\bm{z})d\bm{x}d\bm{z}$. Then, $m_z, \chi_z, q_z$ are approximate values of $\bm{z}_0^\top \hat{\bm{z}}/M$, $\beta(\int \|\bm{z}\|_2^2p(\bm{x},\bm{z})d\bm{x}d\bm{z} - \|\hat{\bm{z}}\|_2^2)/M$, and $\|\hat{\bm{z}}\|_2^2/M$, respectively.

%%%
\subsection{Replica analysis}
% result of the replica analysis
In general, typical values of the macroscopic variables that appear in SE such as $\mathbb{E}_{A,y,x_0}[ ||\hat{\bm{x}}||_2^2/N]$ can be assessed in computing the so-called \emph{free energy} $f=-\lim_{N\to\infty}\mathbb{E}_{A,\bm{y},\bm{x}_0}[\log Z]/(N\beta)$ using the replica method of statistical mechanics \cite{mezard1987spin, mezard2009information, tanaka2002statistical, kabashima2008inference} in the limit of $N,M \to \infty, M/N=\delta \in (0,\infty)$. In the standard RS computation \cite{mezard2009information, mezard1987spin}, this is reduced to an extreme value problem as:
% $-f=g_{\rm F} + g_{\rm G} - g_{\rm S}$ where 
\begin{align}
    f&=-\mathop{\rm extr}_{m_x, \chi_x, q_x, m_z, \chi_z, q_z}[g_{\rm F} + g_{\rm G} - g_{\rm S}],
    \nonumber \\
    g_{\rm F} &=\mathop{\rm extr}_{\hat{m}_{1x}, \hat{\chi}_{1x}, \hat{Q}_{1x}, \hat{m}_{1z}, \hat{\chi}_{1z}, \hat{Q}_{1z}}\left[
            \frac{1}{2}(q_x + \frac{\chi_x}{\beta}) \hat{Q}_{1x} - \frac{1}{2}\chi_x \hat{\chi}_{1x}
        % \right.
        % \nonumber \\
        % &
        % \left.
        -\hat{m}_{1x} m_x
            -\delta \hat{m}_{1z} m_z
        +\frac{\delta}{2}\left(
                (q_z + \frac{\chi_z}{\beta}) \hat{Q}_{1z} - \chi_z \hat{\chi}_{1z}
            \right)
        \right.
        \nonumber \\
        &\left.
            + \int q_{x_0}\hspace{-1pt}(x_0)\phi_x dx_0D\xi_x
            +\delta \int \sqrt{\frac{\hat{T}_{z}}{2\pi}}e^{-\frac{\hat{T}_z}{2}z_0^2}q_{y|z}\hspace{-1pt}(y| z_0) \phi_z dydz_0 D\xi_z
    \hspace{-1pt}\right], 
    \nonumber 
    \\
    g_{\rm G} &= \mathop{\rm extr}_{\hat{m}_{2x}, \hat{\chi}_{2x}, \hat{Q}_{2x}, \hat{m}_{2z}, \hat{\chi}_{2z}, \hat{Q}_{2z}}\left[
            \frac{1}{2}(q_x + \frac{\chi_x}{\beta}) \hat{Q}_{2x} - \frac{1}{2}\chi_x \hat{\chi}_{2x}
    % \right.
    % \nonumber \\
    % &\left.
        - m_x \hat{m}_{2x}- \delta m_z \hat{m}_{2z} + \frac{\delta}{2}
        \left(
            (q_z + \frac{\chi_z}{\beta}) \hat{Q}_{2z} - \chi_z \hat{\chi}_{2z}
        \right)
    \right.
    \nonumber
    \\
    &\left.
        -\frac{1}{2}
        \left\{
            \mathbb{E}_\lambda
            \left[
                \log (\hat{Q}_{2x}+\lambda\hat{Q}_{2z})
            \right]
            -
            \mathbb{E}_\lambda
            \left[
                \frac{
                    \hat{\chi}_{2x} + \lambda \hat{\chi}_{2z}
                }{
                    \hat{Q}_{2x} + \lambda \hat{Q}_{2z}
                }
            \right]
        \right.
    % \right.
    % \nonumber
    % \\
    % &\left.
        \left.
            -\mathbb{E}_\lambda
            \left[
                \frac{
                    T_x(\hat{m}_{2x} + \lambda \hat{m}_{2z})^2
                }{
                    (\hat{Q}_{2x} + \lambda \hat{Q}_{2z})
                }
            \right]
        \right\}
    \right],
    \nonumber 
    \\
    g_{\rm S} &=  \frac{1}{2}
    \hspace{-1pt}\left(\hspace{-1pt}
        \frac{\log \chi_x}{\beta} 
        \hspace{-1pt}+\hspace{-1pt}
        \frac{q_x}{\chi_x} 
        \hspace{-1pt}-\hspace{-1pt} \frac{m_x^2}{T_x \chi_x}
    \hspace{-1pt}\right) \hspace{-1pt}
    \hspace{-1pt}+\hspace{-1pt}
    \frac{\delta}{2}
    \hspace{-1pt}\left(\hspace{-1pt}
        \frac{\log \chi_z}{\beta} 
        \hspace{-1pt}+\hspace{-1pt}
        \frac{q_z}{\chi_z} 
        \hspace{-1pt}-\hspace{-1pt}
        \frac{m_z^2}{T_z\chi_z}
    \hspace{-1pt}\right)\hspace{-1pt}
    \nonumber.
\end{align}
The extreme condition yields the same form of the equations that appear in the fixed point condition of the SE equations (\ref{eq:factorized x 1})-(\ref{eq:message GtoF 4}).
At extremum, the variational parameters $m_x, \chi_x$, and $q_x$ accord with
\begin{equation*}
    \mathbb{E}\left[
        \frac{\bm{x}_0^\top \hat{\bm{x}}}{N}
    \right], \quad
    \mathbb{E}\left[
        \frac{\beta(\int \|\bm{x}\|_2^2p(\bm{x},\bm{z})d\bm{x}d\bm{z} - \|\hat{\bm{x}}\|_2^2)}{N}
    \right], \quad \mbox{and}\quad 
    \mathbb{E}\left[
            \frac{\|\hat{\bm{x}}\|_2^2}{N}
        \right],
\end{equation*}
% $\mathbb{E}[\bm{x}_0^\top \hat{\bm{x}}/N]$, 
% $\mathbb{E}[\beta(\int \|\bm{x}\|_2^2p(\bm{x},\bm{z})d\bm{x}d\bm{z} - \|\hat{\bm{x}}\|_2^2)/N]$, 
% and $\mathbb{E}[\|\hat{\bm{x}}\|_2^2/N]$,
respectively.
Similar accordance also holds between $m_z, \chi_z, q_z$ and
\begin{equation*}
    \mathbb{E}\left[
        \frac{\bm{z}_0^\top \hat{\bm{z}}}{N}
    \right], \quad
    \mathbb{E}\left[
        \frac{\beta(\int \|\bm{x}\|_2^2p(\bm{x},\bm{z})d\bm{x}d\bm{z} - \|\hat{\bm{z}}\|_2^2)}{N}
    \right], \quad \mbox{and}\quad 
    \mathbb{E}\left[
            \frac{\|\hat{\bm{z}}\|_2^2}{N}
        \right].
\end{equation*}
% $\mathbb{E}[\bm{z}_0^\top \hat{\bm{z}}/M]$, 
% $\mathbb{E}[\beta(\int \|\bm{z}\|_2^2p(\bm{x},\bm{z})d\bm{x}d\bm{z} - \|\hat{\bm{z}}\|_2^2)/M]$, $\mathbb{E}[\|\hat{\bm{z}}\|_2^2/M]$.
Thus, the fixed point of VAMP's SE is consistent with the RS calculation. This is the second result of this study. 

% validity condition of above analysis
The aforementioned analysis is valid only when the system is stable against replica symmetry breaking (RSB)\cite{dealmeida1978stability}. The local instability condition of the RS solution against infinitesimal perturbation of the form of the one-step RSB yields:
\begin{align}
    \left(1-2\frac{\partial^2\mathcal{F}(\chi_x,\chi_z)}{\partial \chi_x^2}\chi_x^{(2)}\right)&\left(1 - \frac{2}{\delta}\frac{\partial^2\mathcal{F}(\chi_x,\chi_z)}{\partial \chi_z^2}\chi_z^{(2)}\right) 
    % \nonumber \\
    % &\hspace{-40pt}
    - 
    \frac{4}{\delta}
    \left(
        \frac{\partial^2 \mathcal{F}(\chi_x,\chi_z)}{\partial \chi_x \partial \chi_z}
    \right)
    \chi_x^{(2)}\chi_z^{(2)}<0,
    \label{eq:AT-replica}
\end{align}
where the function $\mathcal{F}(\chi_x, \chi_z)$ is defined as: 
$
    2\mathcal{F}(\chi_x,\chi_z)=\mathop{\rm extr}_{\gamma_x, \gamma_y}
    [
        \chi_x\gamma_x + \delta \chi_z \gamma_y - \mathbb{E}_\lambda
        [
            \log (\gamma_x + \lambda \gamma_y)
        ]
    ]
    -\log \chi_x - \delta \log \chi_z$, and 
    % $\chi_x^{(2)} = \int (\frac{\partial^2 \phi_x}{\partial (\sqrt{\hat{\chi}_{1x}}\xi_{x})^2})^2 q_{x_0}(x_0)dx_0D\xi_x$
    $\chi_x^{(2)} = \int [\partial^2 \phi_x/\partial (\sqrt{\hat{\chi}_{1x}}\xi_{x})^2]^2 q_{x_0}(x_0)dx_0D\xi_x,$
    {}
    $\chi_z^{(2)} = \int [\partial^2 \phi_z/\partial (\sqrt{\hat{\chi}_{1z}}\xi_z)^2]^2 e^{-\frac{\hat{T}_z}{2}z_0^2}q_{y|z}(y|z_0)dydz_0 D\xi_z$.
The equation (\ref{eq:AT-replica}) corresponds to the de Almeida-Thouless (AT) instability condition \cite{kabashima2008inference,dealmeida1978stability} of the current system.

%%%
\subsection{Microscopic instability of VAMP and AT instability}  % microscopic instability of VAMP
% why do we consider microscopic instability condition?, what kind of instability?
The iteration of the equations (\ref{eq:factorized x 1})-(\ref{eq:message GtoF 4}) describes the macroscopic behavior of VAMP.
However, the convergence of macroscopic variables does not directly indicate the convergence of microscopic variables including $\bm{h}_{1x}$ and $ \bm{h}_{1z}$.
Here, we examine whether $\bm{h}_{1x}$ and $\bm{h}_{1z}$ of Algorithm \ref{algo:VAMP} are stable when small perturbations $\epsilon_x \bm{\eta}_{F, x}$ and $\epsilon_z \bm{\eta}_{F, z}$ are added around their fixed points, where each entry of $\bm{\eta}_{F,x}\in\mathbb{R}^N$ and $\bm{\eta}_{F,z}\in\mathbb{R}^M$ is independent random variable of zero mean and unit variance. Let us denote the variables without time indices, for example $\hat{Q}_{2x}$, as the values at the fixed point of Algorithm \ref{algo:VAMP}. Then, linearization around the fixed points indicates that $\epsilon_x$ and $\epsilon_z$ grow exponentially 
by the VAMP iterations, and therefore, the fixed points are unstable if 
\begin{align}
    1 
    &
    -
    \left(
        \frac{1}{\chi_x^2} - \frac{\zeta_2}{\zeta_0\zeta_2 - \zeta_1^2} 
    \right)
    \chi_x^{(2)}
    - 
    \left(
        \frac{1}{\chi_z^2} - \frac{\delta \zeta_0}{\zeta_0 \zeta_2 - \zeta_1^2}
    \right)
    \chi_z^{(2)}
    \nonumber \\
    &
    + \left\{
        \frac{1}{\chi_x^2 \chi_z^2} - \frac{\delta\zeta_0}{\chi_x^2(\zeta_0\zeta_2 - \zeta_1)}
    % \right.
    % \nonumber \\
    % &
    % \hspace{15pt} \left.
        - \frac{\zeta_2}{\chi_z^2(\zeta_0 \zeta_2 - \zeta_1^2)} + \frac{\delta}{\zeta_0 \zeta_2 - \zeta_1^2}
    \right\}
    \chi_x^{(2)}\chi_z^{(2)}
    < 0,
    \label{eq:micro instability}
\end{align}
holds, where $\zeta_0 =\mathbb{E}_\lambda[1/(\hat{Q}_{2x}+\lambda\hat{Q}_{2z})^2], \zeta_1=\mathbb{E}_{\lambda}[\lambda/(\hat{Q}_{2x}+\lambda\hat{Q}_{2z})^2]$, and $\zeta_2 = \mathbb{E}_{\lambda}[\lambda^2/(\hat{Q}_{2x}+\lambda\hat{Q}_{2z})^2]$.

% coincidence with the AT condition derived in the previous section
By the way, the extremum condition of the function $\mathcal{F}$ yields
\begin{align*}
\frac{\partial^2 \mathcal{F}(\chi_x, \chi_z)}{\partial \chi_x^2} &= \frac{1}{2}\left(
        \frac{1}{\chi_x^2} - \frac{\zeta_2}{\zeta_0\zeta_2 - \zeta_1^2}
\right), 
\\
\frac{\partial^2 \mathcal{F}(\chi_x, \chi_z)}{\partial \chi_z^2} &= \frac{\delta}{2}\left(
        \frac{1}{\chi_z^2} - \frac{\delta\zeta_0}{\zeta_0\zeta_2 - \zeta_1^2}
    \right)
\\
\frac{\partial^2 \mathcal{F}(\chi_x, \chi_z)}{\partial \chi_x \partial\chi_z} &= \frac{\delta}{2}\frac{\zeta_1}{\zeta_0\zeta_2 - \zeta_1^2}.
\end{align*}
These indicate that the AT condition (\ref{eq:AT-replica}) agrees with microscopic instability condition (\ref{eq:micro instability}). This means that a spontaneous RSB prevents VAMP from converging, which is the third result of this study. 

%%%%%
\section{Experimental validation} 
\label{section:numerical experiment}

%%%%% figure: time evolution %%%%%
    \begin{figure}[t]
      \centering
      \includegraphics[width= \columnwidth]{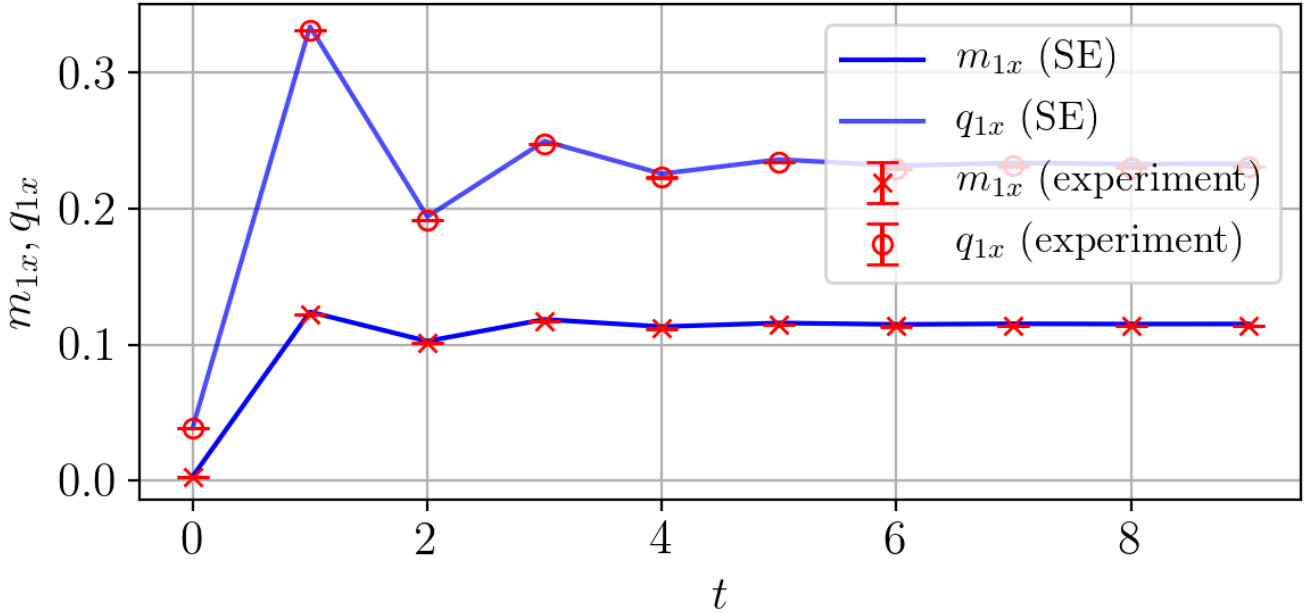}
      \caption{Macroscopic variables $m_{1x}^{(t)}$ and $q_{1x}^{(t)}$ defined in (\ref{eq:factorized x 1}) and (\ref{eq:factorized x 3}) versus algorithm iteration. The solid blue lines are the SE trajectories. The red symbols represent the mean of VAMP trajectories, which are obtained from 8192 experiments. The error bars represent standard errors. 
      }
      \label{fig:state evolution}
    \end{figure}

%%%
\subsection{SE} % validation of our SE
\label{subsection:state evolution}

% experiment setup
To validate the SE, we conducted numerical experiments with the following model mismatch setting. The actual channel and actual prior were specified as the sign function $q_{\bm{y}|\bm{z}}(\bm{y}|\bm{z}) = \delta(\bm{y} - {\rm sign}(\bm{z}))$ and the Bernoulli-Gauss distribution $q_{\bm{x}_0}(\bm{x}_0) = \prod_{i=1}^N(\rho e^{-x_{0,i}^2/2}/\sqrt{2\pi} + (1-\rho)\delta(x_{0,i}))$, respectively. The postulated channel and postulated prior were specified as Gaussian $p_{\bm{y}|\bm{z}}(\bm{y}|\bm{z})\propto\exp(-\|\bm{y}-\bm{z}\|_2^2/2)$ and Laplace distribution $p_{\bm{x}_0}(\bm{x})\propto\exp(-\gamma \|\bm{x}\|_1)$, respectively. Parameter $\beta$ was taken to infinity.  The system size $N$, measurement ratio $\delta=M/N$, and sparsity $\rho$ were specified as $N=1024$, $\delta=0.4$, and $\rho=0.1$, respectively. The measurement matrix $A$ was drawn from the row-orthogonal ensemble\cite{kabashima2014signal}, where the limiting eigenvalue distribution of $A^\top A$ was $\rho(\lambda)=\rho\delta(\lambda -1) +(1-\rho)\delta(\lambda)$. To generate the graphs, we performed $8192$ random trials by forming random measurement matrix $A$.

% result
Figure \ref{fig:state evolution} plots the $m_{1x}^{(t)}$ and $q_{1x}^{(t)}$ defined in (\ref{eq:factorized x 1}) and (\ref{eq:factorized x 3}) versus the number of algorithm iterations. The error bars refer to standard errors. The figure shows that the VAMP trajectories were in excellent accordance with those of the SE. Similar accordance was also obtained for the other macroscopic variables. These show the validity of our SE equations.

%%%
\subsection{AT instability and convergence of VAMP}
\label{subsection:AT instability}

%%%%% figure: convergence probability %%%%%
    \begin{figure}[t]
      \centering
      \includegraphics[width= \columnwidth]{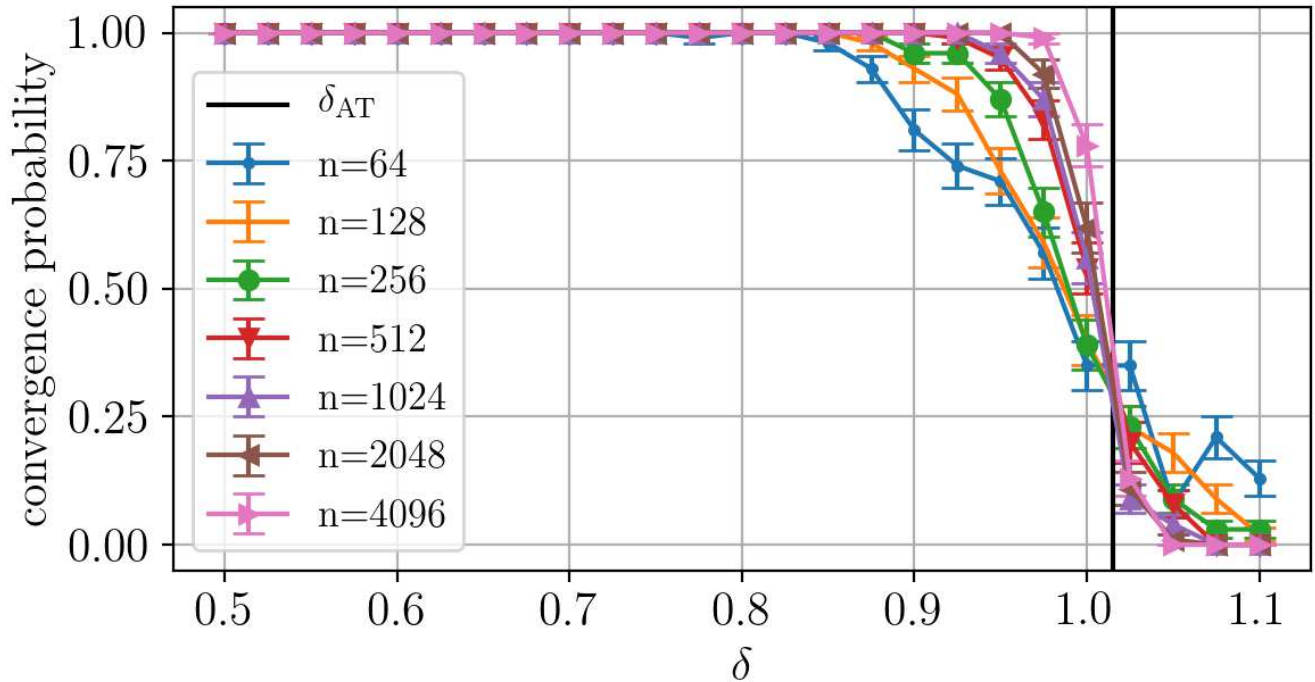}
      \caption{Convergence probability of VAMP versus measurement ratio $\delta$ for the random classification problem. The symbols are the convergence probabilities, which were evaluated from 100 experiments. The solid black line refers to the de Almeida-Thouless instability line reported in \cite{krauth1989storage}.
      }
      \label{fig:convergence}
    \end{figure}
    
% experiment setup
To examine the convergence criteria (\ref{eq:AT-replica}) and (\ref{eq:micro instability}) of VAMP, we conducted a numerical experiment using the random classification problem \cite{krauth1989storage}. In this problem, the actual channel and actual prior were $q_{\bm{y}|\bm{z}}(\bm{y}|\bm{z})\propto\prod_{\mu=1}^M [\delta(y_\mu - 1) + \delta(y_\mu + 1)] $ and $q_{\bm{x}_0}(\bm{x}_0)\propto{\rm Const.}$, respectively. The postulated channel and postulated prior were $p_{\bm{y}|\bm{z}}(\bm{y}|\bm{z})\propto\prod_{\mu=1}^M\Theta(y_\mu z_\mu)$, where $\Theta(x)=1$ if $x>0$, and $0$ otherwise, and Ising prior $p_{\bm{x}_0}(\bm{x})\propto \prod_{i=1}^N [\delta(x_i-1) + \delta(x_i+1)]$, respectively. Here $\beta$ is unity. Each entry of $A$ was drawn from i.i.d. zero-mean Gaussian variables of variance $1/N$. \cite{krauth1989storage} revealed that this system exhibited RSB at a measurement ratio $\delta_{\rm AT}\simeq 1.015$ in the limit $N\to\infty$.

% experiment procedure
To investigate the convergence of VAMP for this problem, we ran VAMP for many instances and calculated convergence probability. The algorithm was considered to have converged when $\|\hat{\bm{x}}_1^{(T_{\rm iter})}-\hat{\bm{x}}_2^{(T_{\rm iter})}\|_2^2/N$ was smaller than $10^{-15}$. The number of iteration was $T_{\rm iter}=10000$.  The system sizes were $N=64$, $128$, $256$, $512$, $1024$, $2048$, and $4096$.

% result
Figure \ref{fig:convergence} plots the convergence probability, which was evaluated from 100 experiments forming random measurement matrix $A$, versus measurement ratio $M/N=\delta$. The figure shows that for large $N$, (i) VAMP converged up close to $\delta_{\rm AT}$ with high probability, (ii) VAMP did not converge above $\delta_{\rm AT}$ with high probability.  The figure demonstrates the coincidence of the instability conditions (\ref{eq:AT-replica}) and (\ref{eq:micro instability}).

%%%%%
\section{Summary and conclusion}
\label{section:summary and conclusion}
% summary and conclusion
In this study, we investigated the behavior of VAMP when the actual and postulated posteriors were mismatched. For rotationally invariant random measurement matrices, we derived the SE equations and showed their fixed points were consistent with the RS solution obtained by the replica method of statistical mechanics. In addition, we showed that the fixed point of VAMP could exhibit a microscopic instability, the critical condition of which agrees with that for breaking the replica symmetry. This correspondence implies that the replica symmetry breaking plays a crucial role in characterizing a fundamental limit of VAMP.

%%%%%%
%% Appendix:
%% If needed a single appendix is created by
%%
%\appendix
%%
%% If several appendices are needed, then the command
%%
\appendices
% \onecolumn
%%
%% in combination with further \section-commands can be used.
%%%%%%
\section{Derivation of SE}
We describe the derivation of the SE equations (\ref{eq:factorized x 1})-(\ref{eq:message GtoF 4}). First, we provide the expression of macroscopic variables (\ref{eq:factorized x 1})-(\ref{eq:factorized z 3}) and (\ref{eq:gaussian x 1})-(\ref{eq:gaussian z 3}) in the limit $N\to\infty$. Then, we derive the update rule for $\hat{m}_{kx}^{(t)}, \hat{m}_{kz}^{(t)},\hat{\chi}_{kx}^{(t)}$, and $\hat{\chi}_{kz}^{(t)}, (k=1,2)$ (\ref{eq:message FtoG 1})-(\ref{eq:message FtoG 4}) and (\ref{eq:message GtoF 1})-(\ref{eq:message GtoF 4}). %Because the derivation of the SE equations for the variable $z$ are similar to those of the variable $x$, we detail the derivation only for the variable $x$.

%%%
\subsection{Derivation of (\ref{eq:factorized x 1})-(\ref{eq:factorized z 3}) and (\ref{eq:gaussian x 1})-(\ref{eq:gaussian z 3})}
For $m_{1x}^{(t)}$, using the definition of $m_{1x}^{(t)}$ and $\hat{\bm{x}}_1^{(t)}$, we can write $m_{1x}^{(t)}$ as 
\begin{align}
    m_{1x}^{(t)} &= \frac{1}{N}\bm{x}_0^\top \hat{\bm{x}}_1^{(t)} 
    \nonumber \\
    &= \frac{1}{N}\sum_{i=1}^N x_{0,i}
    \frac{\partial}{\partial h_{1x,i}^{(t)}}
        \frac{1}{\beta}\log \int e^{
            -\frac{\beta\hat{Q}_{1x}^{(t)}}{2}x_{i}^2 + \beta h_{1x,i}^{(t)}x_i
        }
        p_{x_0}(x_i)^\beta dx_i
    .
    \label{eq:m_1x summation derive}
    % \nonumber 
\end{align}
By the assumption on $\bm{h}_{1x}^{(t)}$, the summation in (\ref{eq:m_1x summation derive}) can be replaced with the expectation:
\begin{align}
    &\frac{1}{N}\sum_{i=1}^N x_{0,i}\frac{\partial}{\partial h_{1x,i}^{(t)}}\frac{1}{\beta}\log \int e^{-\frac{\beta\hat{Q}_{1x}^{(t)}}{2}x_{i}^2 + \beta h_{1x,i}^{(t)}x_i}p_{x_0}(x_i)^\beta dx_i
    \nonumber \\
    &= \int x_0 \frac{\partial}{\partial \hat{m}_{1x}^{(t)}}
    \left[
        \frac{1}{\beta}\log \int e^{-\frac{\beta\hat{Q}_{1x}^{(t)}}{2}x^2}
    % \right.
    % \nonumber \\
    % &
    % \hspace{30pt}\left.
    %     \times
        e^{\beta (\hat{m}_{1x}^{(t)}x_0 + \sqrt{\hat{\chi_{1x}^{(t)}}}\xi_{1x})x}p_{x_0}(x)^\beta dx
    \right]q_0(x_0) dx_0 D\xi_{1x}.
    \nonumber \\
    &= \int x_0 \frac{\partial \phi_x(\hat{m}_{1x}^{(t)}, \hat{Q}_{1x}^{(t)}, \hat{\chi}_{1x}^{(t)};x_0)}{\partial \hat{m}_{1x}^{(t)}} q_0(x_0)dx_0 D\xi_{1x}
    \nonumber
    .
\end{align}
The last equation offers (\ref{eq:factorized x 1}). The equations (\ref{eq:factorized x 2})-(\ref{eq:factorized z 3}) are derived similarly.

For $m_{2x}^{(t)}$, let us denote by $\lambda_i, (i=1,2,\dots ,N)$ the eigenvalues of $A^\top A$. Then, using the singular value decomposition $A=USV^\top$ and the definition of $\hat{\bm{x}}_{2x}^{(t)}$, we can write $m_{2x}^{(t)}$ as 
\begin{align}
    m_{2x}^{(t)} &= \frac{1}{N}(V^\top \bm{x}_0)^\top (\hat{Q}_{2x}^{(t)}I_N + \hat{Q}_{2z}S^\top S)^{-1}
    % \nonumber \\
    % &\hspace{20pt}\times
    (V^\top \bm{h}_{2x}^{(t)} + S^\top U^\top \bm{h}_{2z}^{(t)}).
    \nonumber \\
    &=\frac{1}{N}\sum_{i=1}^N \frac{(V^\top \bm{x}_0)_i(V^\top \bm{h}_{2x}^{(t)} + S^\top U^\top \bm{h}_{2z}^{(t)})_i}{\hat{Q}_{2x}^{(t)} + \lambda_i\hat{Q}_{2z}^{(t)}}.
    \label{eq:m_2x summation}
\end{align}
Using the assumption, the equation (\ref{eq:m_2x summation}) can be rewritten as follows:
\begin{align}
    &\frac{1}{N}\sum_{i=1}^N \frac{(V^\top \bm{x}_0)_i(V^\top \bm{h}_{2x}^{(t)} + S^\top U^\top \bm{h}_{2z}^{(t)})_i}{\hat{Q}_{2x}^{(t)} + \lambda_i\hat{Q}_{2z}^{(t)}}
    \nonumber \\
    &\simeq \frac{1}{N}\sum_{i=1}^N \frac{
            (V^\top \bm{x}_0)_i^2 (\hat{m}_{2x}^{(t)} + \lambda_i \hat{m}_{2z}^{(t)})
        }
        {
            \hat{Q}_{2x}^{(t)} + \lambda_i\hat{Q}_{2z}^{(t)}
        }
    % \nonumber \\
    %     &
        +
        \frac{1}{N}\sum_{i=1}^N \frac{
             (V^\top\bm{x}_0)_i(
                \sqrt{\hat{\chi}_{2x}^{(t)}}\xi_{2x,i}^{(t)} 
                + \sqrt{\hat{\chi}_{2z}^{(t)}}(S^\top \bm{\xi}_{2z}^{(t)})_i
            )
        }{
            \hat{Q}_{2x}^{(t)} + \lambda_i\hat{Q}_{2z}^{(t)}
        }
    \nonumber 
    \\
    &\xrightarrow{N\to\infty} \frac{1}{N}\|\bm{x}_0\|_2^2 \mathbb{E}_{\lambda} \left[\frac{\hat{m}_{2x}^{(t)} + \lambda \hat{m}_{2z}^{(t)}}{\hat{Q}_{2x}^{(t)} + \lambda \hat{Q}_{2z}^{(t)}}\right]
    \nonumber.
\end{align}
Because $V$ is assumed to be generated from the uniform distribution on $N\times N$ orthogonal matrices, $V^\top \bm{x}_0$ behaves like a zero-mean Gaussian variable with variance $\|\bm{x}_0\|_2^2/N$. This property and the independence of $\bm{\xi}_{2x}^{(t)}$ and $\bm{\xi}_{2z}^{(t)}$ offeres the last expression. The equations (\ref{eq:gaussian x 2})-(\ref{eq:gaussian z 3}) are obtained similarly.

%%%
\subsection{Derivation of (\ref{eq:message FtoG 1})-(\ref{eq:message FtoG 4}) and (\ref{eq:message GtoF 1})-(\ref{eq:message GtoF 4})}
From the assumption, $\hat{m}_{kx}^{(t)}$ and $\hat{\chi}_{kx}^{(t)}$, $(k=1,2)$ can be obtained as 
\begin{align}
    \hat{m}_{kx}^{(t)} &= \frac{1}{\|\bm{x}_0\|_2^2}\bm{x}_0^\top \bm{h}_{kx}^{(t)}, 
    \nonumber \\
    \hat{\chi}_{kx}^{(t)} &= \frac{1}{N}\|\bm{h}_{kx}^{(t)} - \hat{m}_{kx}^{(t)}\bm{x}_0\|_2^2,
    \nonumber 
\end{align}
For $\hat{m}_{2x}^{(t)}$, the update rule of VAMP and the definition of $m_{1x}^{(t)}$ directly offers the following result:
\begin{align}
    \hat{m}_{2x}^{(t)} &= \frac{1}{\|\bm{x}_0\|_2^2}\bm{x}_0^\top\left(
        \frac{\hat{\bm{x}}_1^{(t)}}{\chi_{1x}^{(t)}} - \bm{h}_{1x}^{(t)}
    \right)
    \nonumber
     \\
    &= \frac{\bm{x}_0^\top \hat{\bm{x}}_1^{(t)}}{\|\bm{x}_0\|_2^2\chi_{1x}^{(t)}} - \frac{\bm{x}_0^\top \bm{h}_{1x}^{(t)}}{\|\bm{x}_0\|_2^2}
    \nonumber \\
    &= \frac{m_{1x}^{(t)}}{T_x\chi_{1x}^{(t)}} - \hat{m}_{1x}^{(t)}.
    \label{eq:m_hat}
\end{align}
This is the equation (\ref{eq:message FtoG 1}). The update rules for $\hat{m}_{1x}^{(t)}$ can be derived in the same way.

For $\hat{\chi}_{2x}^{(t)}$, its update rule can be derived as follows:
\begin{align}
    \hat{\chi}_{2x}^{(t)} &\overset{\rm (a)}{=}  \frac{1}{N}\left\|
        \frac{\hat{\bm{x}}_1^{(t)}}{\chi_{1x}^{(t)}} - \frac{m_{1x}^{(t)}}{T_x\chi_{1x}^{(t)}}\bm{x}_0 - (\bm{h}_{1x}^{(t)} - \hat{m}_{1x}^{(t)}\bm{x}_0)
    \right\|_2^2
    \nonumber \\
    &\overset{\rm (b)}{=} \frac{q_{1x}^{(t)}}{(\chi_{1x}^{(t)})^2} - \frac{(m_{1x}^{(t)})^2}{T_x(\chi_{1x}^{(t)})^2} + \hat{\chi}_{1x}^{(t)}
    % \nonumber \\
    +\frac{2m_{1x}^{(t)}}{T_x\chi_{1x}^{(t)}}\frac{(\bm{h}_{1x}^{(t)} - \hat{m}_{1x}^{(t)}\bm{x}_0)^\top \bm{x}_0}{N} - \frac{2}{\chi_{1x}^{(t)}} \frac{(\bm{h}_{1x}^{(t)} - \hat{m}_{1x}^{(t)}\bm{x}_0)^\top \hat{\bm{x}}_1^{(t)}}{N}
    \nonumber \\
    &\xrightarrow{\rm (c)} \frac{q_{1x}^{(t)}}{(\chi_{1x}^{(t)})^2} - \frac{(m_{1x}^{(t)})^2}{T_x(\chi_{1x}^{(t)})^2} - \hat{\chi}_{1x}^{(t)}, (N\to\infty)
    \nonumber,
\end{align}
where the equality (a) follows from the update rule of VAMP and the SE for $\hat{m}_{2x}^{(t)}$; (b) follows from the definition of $q_{1x}^{(t)}, m_{1x}^{(t)}$ and $\hat{\chi}_{1x}^{(t)}$; the limit (c) can be obtained from the assumption of $\bm{h}_{1x}^{(t)}$, the definition of $\hat{\bm{x}}_1^{(t)}$ and the identity $\int xf(x)Dx = \int f^\prime (x) Dx$
for $\forall{f}(x)$:
\begin{align}
    \frac{(\bm{h}_{1x}^{(t)} - \hat{m}_{1x}^{(t)}\bm{x}_0)^\top \bm{x}_0}{N}
    % \nonumber \\
    % &
    &\simeq \frac{1}{N}\sum_{i=1}^N\sqrt{\hat{\chi}_{1x}^{(t)}}\xi_{1x,i}^{(t)} x_{0,i}
    \nonumber \\
    &\xrightarrow{N\to\infty} \sqrt{\hat{\chi}_{1x}^{(t)}}\int \xi_{1x}x_0q_{x_0}(x_0)dx_0D\xi_{1x} = 0
    \nonumber, \\
    \frac{(\bm{h}_{1x}^{(t)} - \hat{m}_{1x}^{(t)}\bm{x}_0)^\top \hat{\bm{x}}_1^{(t)}}{N}
    % \nonumber \\
    &\simeq \frac{1}{N}\sum_{i=1}^N\sqrt{\hat{\chi}_{1x}^{(t)}} \xi_{1x,i}^{(t)} \frac{\partial }{\partial (\sqrt{\hat{\chi}_{1x}^{(t)}}\xi_{1x,i}^{(t)})}\left[
        \frac{1}{\beta}\log \int e^{-\frac{\beta\hat{Q}_{1x}^{(t)}}{2}x_i^2}
    % \right.
    % \nonumber \\
    % &\hspace{20pt}\left.
    %     \times
        e^{\beta (\hat{m}_{1x}^{(t)}x_{0,i} + \sqrt{\hat{\chi}_{1x}^{(t)}}\xi_{1x,i}^{(t)})}p_{x_0}(x_i)^\beta dx_i
    \right]
    \nonumber \\
    \xrightarrow{N\to\infty}\int& \sqrt{\hat{\chi}_{1x}^{(t)}} \xi_{1x}\frac{\partial }{\partial (\sqrt{\hat{\chi}_{1x}^{(t)}}\xi_{1x})}\left[
        \frac{1}{\beta}\log \int e^{-\frac{\beta\hat{Q}_{1x}^{(t)}}{2}x^2}
    % \right.
    % \nonumber \\
    % &\hspace{20pt}\left.
    %     \times 
        e^{\beta (\hat{m}_{1x}^{(t)}x_{0} + \sqrt{\hat{\chi}_{1x}^{(t)}}\xi_{1x})}p_{x_0}(x)^\beta dx
    \right] q_{x_0}(x_0)dx_0 D\xi_{1x}
    \nonumber \\
    &=\hat{\chi}_{1x}^{(t)}\int \frac{\partial^2 }{\partial (\sqrt{\hat{\chi}_{1x}^{(t)}}\xi_{1x})^2}\left[
        \frac{1}{\beta}\log \int e^{-\frac{\beta\hat{Q}_{1x}^{(t)}}{2}x^2}
    % \right.
    % \nonumber \\
    % &\hspace{20pt}\left.
    %     \times 
        e^{\beta (\hat{m}_{1x}^{(t)}x_{0} + \sqrt{\hat{\chi}_{1x}^{(t)}}\xi_{1x})}p_{x_0}(x)^\beta dx
    \right] q_{x_0}(x_0)dx_0 D\xi_{1x}
    \nonumber \\
    &= \hat{\chi}_{1x}^{(t)}\chi_{1x}^{(t)}.
    \nonumber
\end{align}
Analogously, the update rule for $\hat{\chi}_{1x}^{(t+1)}$ is derived as follows:
\begin{align}
    \hat{\chi}_{1x}^{(t+1)} &= \frac{1}{N}\left\|
        \frac{\hat{\bm{x}}_2^{(t)}}{\chi_{2x}^{(t)}} - \frac{m_{2x}^{(t)}}{T_x\chi_{2x}^{(t)}} - (\bm{h}_{2x}^{(t)} - \hat{m}_{2x}^{(t)}\bm{x}_0)
    \right\|_2^2
    \nonumber \\
    &\overset{\rm (a)}{=} \frac{q_{2x}^{(t)}}{(\chi_{2x}^{(t)})^2} - \frac{(m_{2x}^{(t)})^2}{T_x (\chi_{1x}^{(t)})^2} + \hat{\chi}_{2x}^{(t)} 
    % \nonumber \\
    % &
    +\frac{2m_{2x}^{(t)}}{T_x\chi_{2x}^{(t)}}\frac{(\bm{h}_{2x}^{(t)})^\top \bm{x}_0 - \hat{m}_{2x}^{(t)}\|\bm{x}_0\|_2^2}{N}
    % \nonumber \\
    % &
    -\frac{2}{\chi_{2x}^{(t)}}\frac{[V^\top (\bm{h}_{2x}^{(t)} - \hat{m}_{2x}^{(t)}\bm{x}_0)]^\top [V^\top \hat{\bm{x}}_2^{(t)}]}{N}
    \nonumber \\
    &\overset{\rm (b)}{=}\frac{q_{2x}^{(t)}}{(\chi_{2x}^{(t)})^2} - \frac{(m_{2x}^{(t)})^2}{T_x (\chi_{1x}^{(t)})^2} + \hat{\chi}_{2x}^{(t)} 
    % \nonumber \\
    % &
    -\frac{2}{\chi_{2x}^{(t)}}\frac{[V^\top (\bm{h}_{2x}^{(t)} - \hat{m}_{2x}^{(t)}\bm{x}_0)]^\top [V^\top \hat{\bm{x}}_2^{(t)}]}{N}
    \nonumber \\
    &\xrightarrow{\rm (c)} \frac{q_{2x}^{(t)}}{(\chi_{2x}^{(t)})^2} -\frac{(m_{2x}^{(t)})^2}{T_x(\chi_{2x}^{(t)})^2} - \hat{\chi}_{2x}^{(t)}, (N\to\infty),
    \nonumber
\end{align}
where (a) follows from the definition of $q_{2x}^{(t)}, m_{2x}^{(t)}$ and $\hat{\chi}_{2x}^{(t)}$; (b) follows from the definition of $\hat{m}_{2x}^{(t)}$; the limit (c) follows from the assumption on $\bm{h}_{2x}^{(t)}$ and the definition of $\hat{\bm{x}}_{2}^{(t)}$:
\begin{align}
    \frac{[V^\top (\bm{h}_{2x}^{(t)} - \hat{m}_{2x}^{(t)}\bm{x}_0)]^\top [V^\top \hat{\bm{x}}_2^{(t)}]}{N}
    % \nonumber \\
    % &
    &\simeq \frac{1}{N}\sum_{i=1}^N \sqrt{\hat{\chi}_{2x}^{(t)}}\xi_{2x,i}^{(t)}
    \left[
        \frac{
            (\hat{m}_{2x}^{(t)} + \lambda_i\hat{m}_{2z}^{t})(V^\top\bm{x}_0)_i
        }{
            \hat{Q}_{2x}^{(t)} + \lambda_i\hat{Q}_{2z}
        }
        % \nonumber\\
        % &
        % +\frac{1}{N}\sum_{i=1}^N
        +
        \frac{
            \sqrt{\hat{\chi}_{2x}^{(t)}}\xi_{2x,i}^{(t)} +\sqrt{\hat{\chi}_{2z}^{(t)}}(S^\top \bm{\xi}_{2z}^{(t)})_i
        }{
            \hat{Q}_{2x}^{(t)} + \lambda_i\hat{Q}_{2z}
        }
    \right]
    \nonumber \\
    \xrightarrow{N\to\infty}& \sqrt{\hat{\chi}_{2x}^{(t)}} \int \xi_{2x}\tilde{x}_0D\tilde{x}_0 D\xi_{2x} \mathbb{E}_\lambda\left[
        \frac{
            \hat{m}_{2x}^{(t)} + \lambda\hat{m}_{2z}^{t}
        }{
            \hat{Q}_{2x}^{(t)} + \lambda\hat{Q}_{2z}
        }
    \right]
    \nonumber \\
    &
    + \hat{\chi}_{2x}^{(t)}\int \xi_{2x}^2 D\xi_{2x}\mathbb{E}_\lambda\left[
        \frac{1}{\hat{Q}_{2x}^{(t)} + \lambda\hat{Q}_{2z}^{(t)}}
    \right]
    \nonumber \\
    &
    + \sqrt{\hat{\chi}_{2x}^{(t)}\hat{\chi}_{2z}^{(t)}}\int \xi_{2x}\xi_{2z}D\xi_{2x}D\xi_{2z}
    \mathbb{E}_\lambda\left[
        \frac{\sqrt{\lambda}}{\hat{Q}_{2x}^{(t)} + \lambda\hat{Q}_{2z}^{(t)}},
    \right]
    \nonumber \\
    &=\hat{\chi}_{2x}^{(t)}\chi_{2x}^{(t)}.
    \nonumber
\end{align}
The update rule for $\hat{\chi}_{kz}, (k=1,2)$ can be derived in the same way. Furthermore, the update rule for $\hat{Q}_{kx}, \hat{Q}_{kz}, (k=1,2)$ are exactly same with the update rule of VAMP itself.

%%%%%
\section{RS calculation of the free energy}
In this section, we outline the RS calculation of the free energy. Because analogous calculations can be found in \cite{kabashima2008inference, shinzato2008perceptron} and \cite{shinzato2008learning}, we only show the main steps. For a general introduction to the replica method, we refer to \cite{mezard1987spin} and \cite{mezard2009information}. 

In general, the evaluation of the free energy  $f=-\lim_{N\to\infty}\mathbb{E}_{A,\bm{y}, \bm{x}_0}[\log Z]/N\beta$ is technically difficult because it requires the average of the logarithm. To carry out the calculation of the free energy, the replica method of statistical mechanics first rewrites the free energy using an identity $\mathbb{E}[\log Z]=\lim_{n\to 0}n^{-1}\log\mathbb{E}[Z^n]$ as 
\begin{align}
    \begin{split}
    f &= - \lim_{n\to 0}\frac{1}{n}\phi_n,
    \\
    \phi_n &= \lim_{N\to\infty} \frac{1}{N\beta}\log\mathbb{E}_{A,\bm{y},\bm{x}_0}[Z^n].
    \end{split}
    \label{eq:replica identity}
\end{align}
Although the evaluation of $\phi_n$ for $n\in\mathbb{R}$ in a rigorous manner is difficult, this expression has an advantage. For natural number $n=1,2,\ldots$, let us denote by $d^n\bm{x}=d\bm{x}_1\dots d\bm{x}_n$ a measure over $\mathbb{R}^{N\times n}$, with $\bm{x}_1,\dots ,\bm{x}_n\in\mathbb{R}^N$. Analogously, $d^n\bm{z}=d\bm{z}_1\dots d\bm{z}_n$ a measure over $\mathbb{R}^{M\times n}$, with $\bm{z}_1,\dots ,\bm{z}_n\in\mathbb{R}^M$. Then, for $n=1,2,\dots $, using the identity
\begin{align}
    Z^n &= \left(
        \int
            p_{\bm{y}|\bm{z}}(\bm{y}|\bm{z})^\beta p_{\bm{x}_0}(\bm{x})^\beta \delta(A\bm{x} - \bm{z})
        d\bm{x}d\bm{z}
    \right)^n
    \nonumber \\
    &= \int 
        \prod_{a=1}^n
        p_{\bm{y}|\bm{z}}(\bm{y}|\bm{z}_a)^\beta  p_{\bm{x}_0}(\bm{x}_a)^\beta
        \delta(A\bm{x}_a - \bm{z}_a)
    d^n\bm{x}d^n\bm{z}
    \nonumber,
\end{align}
$\phi_n$ can be written as 
\begin{align}
    \phi_n = \lim_{N\to\infty}\frac{1}{N\beta}\log 
            \int 
                % \left\{
                    % \int 
                        % q_{\bm{y}|\bm{z}}(\bm{y}|\bm{z}_0)
                        \prod_{a=1}^n 
                        p_{\bm{y}|\bm{z}}(\bm{y}|\bm{z}_a)^\beta
                    % d\bm{y}d\bm{z}_0
                % \right\}
                % \nonumber \\
                % \times 
                % \left\{
                    % \int 
                        % q_{\bm{x}_0}(\bm{x}_0)
                        % \prod_{a=1}^n
                        p_{\bm{x}_0}(\bm{x}_a)^\beta 
                    % d\bm{x}_0
                % \right\}
                % \nonumber \\
                % \times 
                \mathbb{E}_{A}\left[\prod_{a=0}^n \delta(A\bm{x}_a- \bm{z}_a)\right]
        q_{\bm{y}|\bm{z}}(\bm{y}|\bm{z}_0)q_{\bm{x}_0}(\bm{x}_0)d^n\bm{x}d^n\bm{z}d\bm{x}_0d\bm{z}_0d\bm{y},
        \label{eq:phi_n (general)}
\end{align}
which is much easier to evaluate than the average of the logarithm. The replica method evaluates a formal expression of $\phi_n$ for $n=1,2,\dots $, and then extrapolates it as $n\to0$.

% %%%
% \subsection{Average over $A$}
At this point, we can take the average with respect to $A$ in (\ref{eq:phi_n (general)}) using the singular value decomposition of $A=USV^\top$ and an identity $\prod_{a=0}^n\delta(A\bm{x}_a-\bm{z}_a)=\lim_{\gamma\to\infty}(\gamma/2\pi)^{M(n+1)/2}\prod_{a=0}^n\exp(-\frac{\gamma}{2}\|A\bm{x}_a-\bm{z}_a\|_2^2)$:
\begin{align}
    &\mathbb{E}_{A}\left[\prod_{a=0}^n \delta(A\bm{x}_a- \bm{z}_a)\right]
    \nonumber \\
    &=\lim_{\gamma\to\infty}\left(\frac{\gamma}{2\pi}\right)^{\frac{M(n+1)}{2}}\mathbb{E}_{U,V}\left[
        \prod_{a=0}^n\exp
        \left(
            -\frac{\gamma}{2}(U^\top \bm{z}_a)^\top (U^\top \bm{z}_a)
        % \right.
    % \right.
    %     \nonumber \\
    % &\left.
        % \left.
        + \gamma (U^\top\bm{z}_a)^\top S(V^\top \bm{x}_a)  - \frac{\gamma}{2}(V^\top \bm{x}_a)^\top S^\top S(V^\top \bm{x}_a)
        \right)
    \right].
    \label{eq:average_a}
\end{align}
Because $U$ and $V$ are assumed to be drawn from the uniform distribution over $N\times N$ and $M\times M$ orthogonal matrices, for fixed variable $\bm{x}_a$ and $\bm{z}_a$, $\tilde{\bm{x}}_a = V^\top \bm{x}_a$ and $\tilde{\bm{z}}_a = U^\top \bm{z}_a$ behave as continuous random variables which are uniformly distributed over the constraints
\begin{align}
    \begin{cases}
        \tilde{\bm{x}}_a^\top\tilde{\bm{x}}_b = \bm{x}_a^\top \bm{x}_b
        \nonumber \\
        \tilde{\bm{z}}_a^\top\tilde{\bm{z}}_b = \bm{z}_a^\top \bm{z}_b
        \nonumber 
    \end{cases},
    \quad a,b=0,1,\dots ,n.
\end{align}
Let us denote by $d^{n+1}\tilde{\bm{x}}=d\tilde{\bm{x}}_0d\tilde{\bm{x}}_1\dots d\tilde{\bm{x}}_n$ a measure over $\mathbb{R}^{N\times (n+1)}$, with $\tilde{\bm{x}}_0, \tilde{\bm{x}}_1,\dots ,\tilde{\bm{x}}_n\in\mathbb{R}^N$. Analogously, $d^{n+1}\bm{z}=d\tilde{\bm{z}}_0d\tilde{\bm{z}}_1\dots d\tilde{\bm{z}}_n$ a measure over $\mathbb{R}^{M\times (n+1)}$, with $\tilde{\bm{z}}_0, \tilde{\bm{z}}_1,\dots ,\tilde{\bm{z}}_n\in\mathbb{R}^M$.
Then, by inserting trivial identities
\begin{align}
    1 &= \prod_{0\le a\le b \le n}N\int \delta\left(
            NQ_{x}^{(ab)} - \bm{x}_a^\top \bm{x}_b
        \right)dQ_{x}^{(ab)}, 
    \nonumber \\
    1 &= \prod_{0\le a\le b \le n}M\int \delta\left(
        MQ_{z}^{(ab)} - \bm{z}_a^\top \bm{z}_b
    \right)dQ_{x}^{(ab)}, 
    \nonumber 
\end{align}
into the integrand in (\ref{eq:phi_n (general)}), the equation (\ref{eq:average_a}) can be rewritten in the limit $N,M\to\infty$, with $M/N\to\delta\in(0,\infty)$:
\begin{align}
    \lim_{N\to\infty}\frac{1}{N}\mathbb{E}_A\left[\prod_{a=0}^n \delta(A\bm{x}_a- \bm{z}_a)\right] 
    = \tilde{g}_{\rm G}(Q_x, Q_z) - \tilde{g}_{\rm S}(Q_x, Q_z),
    \label{eq:energy_term}
\end{align}
where
\begin{align*}
    \tilde{g}_{\rm G}(Q_x, Q_z) &= \lim_{N, \gamma\to\infty}\frac{1}{N}\log\left(\frac{\gamma}{2\pi}\right)^{\frac{M(n+1)}{2}}
        \int \prod_{a=0}^ne^{
            -\frac{\gamma}{2}\tilde{\bm{z}}_a^\top \tilde{\bm{z}}_a
        + \gamma \tilde{\bm{z}}_a^\top S\tilde{\bm{x}}_a  - \frac{\gamma}{2}\tilde{\bm{x}}_a^\top S^\top S\tilde{\bm{x}}_a
        }
        \nonumber \\
        &\hspace{120pt}\times
        \prod_{0\le a\le b \le n}\delta(NQ_x^{(ab)} - \tilde{\bm{x}}_a^\top \tilde{\bm{x}}_b )\delta(MQ_z^{(ab)} - \tilde{\bm{z}}_a^\top\tilde{\bm{z}}_b)d^{n+1}\tilde{\bm{x}}d^{n+1}\tilde{\bm{z}} \\
    \tilde{g}_{\rm S}(Q_x,Q_z) &= \lim_{N\to\infty}\frac{1}{N}\log
        \int \prod_{0\le a\le b \le n}\delta(NQ_x^{(ab)} - \tilde{\bm{x}}_a^\top \tilde{\bm{x}}_b )\delta(MQ_z^{(ab)} - \tilde{\bm{z}}_a^\top\tilde{\bm{z}}_b)d^{n+1}\tilde{\bm{x}}d^{n+1}\tilde{\bm{z}}.
\end{align*}
To evaluate $\tilde{g}_{\rm G}(Q_x, Q_z)$ and $\tilde{g}_{\rm S}(Q_x, Q_z)$, the Fourier transform representations of the delta functions are useful:
\begin{align}
    \prod_{0\le a \le b \le n}\delta(NQ_x^{(ab)}-\tilde{\bm{x}}_a^\top\tilde{\bm{x}}_b) &= \prod_{0\le a \le b \le n}\int_{-\sqrt{-1}\infty}^{\sqrt{-1}\infty}e^{\frac{\Lambda_x^{(ab)}}{2}(NQ_x^{(ab)}-\tilde{\bm{x}}_a^\top\tilde{\bm{x}}_b)}\frac{d\Lambda_x^{(ab)}}{4\pi}
    \nonumber \\
    &=\int e^{\frac{N}{2}{\rm Tr}(Q_x\Lambda_x)}\prod_{i=1}^Ne^{-\frac{1}{2}\tilde{\bm{x}}_i^\top \Lambda_x\tilde{\bm{x}}_i}\frac{d\Lambda_x}{(4\pi)^{n(n+1)}}
    \nonumber \\
    \prod_{0\le a \le b \le n}\delta(MQ_z^{(ab)}-\tilde{\bm{z}}_a^\top\tilde{\bm{z}}_b) &= \prod_{0\le a \le b \le n}\int_{-\sqrt{-1}\infty}^{\sqrt{-1}\infty}e^{\frac{\Lambda_z^{(ab)}}{2}(NQ_z^{(ab)}-\tilde{\bm{z}}_a^\top\tilde{\bm{z}}_b)}\frac{d\Lambda_z^{(ab)}}{4\pi}
    \nonumber \\
    &= \int e^{\frac{M}{2}{\rm Tr}(Q_z\Lambda_z)}\prod_{\mu=1}^Me^{-\frac{1}{2}\tilde{\bm{z}}_\mu^\top \Lambda_z\tilde{\bm{z}}_\mu}\frac{d\Lambda_z}{(4\pi)^{n(n+1)}},
    \nonumber
\end{align}
where we denote $\Lambda_x=[\Lambda_x^{(ab)}]$, $\Lambda_z=[\Lambda_z^{(ab)}]$, $Q_x=[Q_x^{(ab)}]$, $Q_z=[Q_z^{(ab)}]\in\mathbb{R}^{(n+1)\times(n+1)}$,  $\tilde{\bm{x}}_i=(\tilde{x}_i^{(0)}, \dots ,\tilde{x}_i^{(n)})\in\mathbb{R}^{n+1}, i=1,2,\dots ,N$, $\tilde{\bm{z}}_\mu=(\tilde{z}_\mu^{(0)}, \dots ,\tilde{z}_\mu^{(n)})\in\mathbb{R}^{n+1}, \mu=1,2,\dots ,M$, $d\Lambda_x=\prod_{a\le b}d\Lambda_x^{(ab)}$, and $d\Lambda_z=\prod_{a\le b}d\Lambda_z^{(ab)}$.
These Fourier transform representations and the saddle point method allow us to evaluate $\tilde{g}_{\rm G}$ as follows:
\begin{align}
    &\tilde{g}_{\rm G}(Q_x,Q_z) = \lim_{N,\gamma\to\infty}\frac{1}{N}\log\left(\frac{\gamma}{2\pi}\right)^{\frac{M(n+1)}{2}}\int \prod_{a=0}^ne^{
            -\frac{\gamma}{2}\tilde{\bm{z}}_a^\top \tilde{\bm{z}}_a
        + \gamma \tilde{\bm{z}}_a^\top S\tilde{\bm{x}}_a  - \frac{\gamma}{2}\tilde{\bm{x}}_a^\top S^\top S\tilde{\bm{x}}_a
        }
        \nonumber \\
        &\hspace{50pt}\times e^{\frac{N}{2}{\rm Tr}(Q_x\Lambda_x)+\frac{M}{2}{\rm Tr}(Q_z\Lambda_z)}\prod_{i=1}^Ne^{-\frac{1}{2}\tilde{\bm{x}}_i^\top \Lambda_x\tilde{\bm{x}}_i}\prod_{\mu=1}^Me^{-\frac{1}{2}\tilde{\bm{z}}_\mu^\top \Lambda_z\tilde{\bm{z}}_\mu}\frac{d\Lambda_x}{(4\pi)^{n(n+1)}}\frac{d\Lambda_z}{(4\pi)^{n(n+1)}}d^{n+1}\tilde{\bm{x}}d^{n+1}\tilde{\bm{z}}
    \nonumber \\
    &= \lim_{N,\gamma\to\infty}\frac{1}{N}\log\left(\frac{\gamma}{2\pi}\right)^{\frac{N\delta(n+1)}{2}} \int \exp\left(N
        \left[
            \frac{1}{2}{\rm Tr}(Q_x \Lambda_x) + \frac{\delta}{2}{\rm Tr}(Q_z \Lambda_z)
            -\frac{\delta-1}{2}\log \det (\Lambda_z + \gamma I_{n+1})
        \right.
    \right.
    \nonumber \\
    &\hspace{150pt}\left.
        \left.
            -\frac{1}{2}\mathbb{E}_\lambda
            [
                \log \det 
                (
                    \Lambda_x\Lambda_z + \gamma (\Lambda_x + \lambda \Lambda_z)
                )
            ]
        \right]
    \right)
    \frac{d\Lambda_x}{(4\pi)^{n(n+1)}}d^{n+1}\frac{d\Lambda_z}{(4\pi)^{n(n+1)}}
    \nonumber \\
    &=\lim_{\gamma\to\infty}
    \left[
        \frac{\delta(n+1)}{2}\log \gamma
        + \mathop{\rm extr}_{\Lambda_x, \Lambda_z}
        \left[
                \frac{1}{2}{\rm Tr}(Q_x \Lambda_x) + \frac{\delta}{2}{\rm Tr}(Q_z \Lambda_z)
            -\frac{\delta - 1}{2}\log \det (\Lambda_z + \gamma I_{n+1})
        \right.
    \right.
    \nonumber \\
    &\hspace{200pt}\left.
        \left.
                -\frac{1}{2}\mathbb{E}_\lambda
                \left[
                    \log \det 
                    \left(
                        \Lambda_x\Lambda_z + \gamma (\Lambda_x + \lambda \Lambda_z)
                    \right)
                \right]
        \right]
    \right]
    \nonumber \\
    &=\mathop{\rm extr}_{\Lambda_x, \Lambda_z}
        \left[
            \frac{1}{2}{\rm Tr}(Q_x \Lambda_x) + \frac{\delta}{2}{\rm Tr}(Q_z \Lambda_z) -\frac{1}{2}\mathbb{E}_\lambda
            \left[
                \log \det 
                \left(
                    \Lambda_x + \lambda \Lambda_z
                \right)
            \right]
        \right].
        % \label{eq:g_tilde_G} 
        \nonumber 
\end{align}
Analogously, $\tilde{g}_{\rm S}$ can be evaluated as 
\begin{align}
    &\tilde{g}_{\rm S}(Q_x, Q_z) 
    \nonumber \\
    &= \lim_{N\to\infty}\frac{1}{N} \log\int e^{\frac{N}{2}{\rm Tr}(Q_x\Lambda_x)+\frac{M}{2}{\rm Tr}(Q_z\Lambda_z)}\prod_{i=1}^Ne^{-\frac{1}{2}\tilde{\bm{x}}_i^\top \Lambda_x\tilde{\bm{x}}_i}\prod_{\mu=1}^Me^{-\frac{1}{2}\tilde{\bm{z}}_\mu^\top \Lambda_z\tilde{\bm{z}}_\mu}\frac{d\Lambda_x}{(4\pi)^{n(n+1)}}\frac{d\Lambda_z}{(4\pi)^{n(n+1)}}d^{n+1}\tilde{\bm{x}}d^{n+1}\tilde{\bm{z}} 
    \nonumber \\
    &=\lim_{N\to\infty}\frac{1}{N}\log\int 
        \exp\left(
            \frac{N}{2}{\rm Tr}(Q_x \Lambda_x) + \frac{M}{2}{\rm Tr}(Q_z \Lambda_z) 
            - \frac{N}{2}\log \det \Lambda_x - \frac{M}{2}\log \det \Lambda_z
        \right)
    \frac{d\Lambda_x}{(4\pi)^{n(n+1)}}\frac{d\Lambda_z}{(4\pi)^{n(n+1)}}
    \nonumber \\
    &= \lim_{N\to\infty}\frac{1}{N}\log\int 
        \exp\left(N
            \left[
                \frac{1}{2}{\rm Tr}(Q_x \Lambda_x) + \frac{\delta}{2}{\rm Tr}(Q_z \Lambda_z) 
            - \frac{1}{2}\log \det \Lambda_x - \frac{\delta}{2}\log \det \Lambda_z
            \right]
        \right)
    \frac{d\Lambda_x}{(4\pi)^{n(n+1)}}\frac{d\Lambda_z}{(4\pi)^{n(n+1)}}
    \nonumber \\
    &= \frac{1}{2}\log \det Q_x + \frac{\delta}{2}\log\det Q_z.
    % \label{eq:g_tilde_S}
    \nonumber 
\end{align}
Here we omit the constants including $\log 2\pi$.

The equation (\ref{eq:energy_term}) indicates that $\phi_n$ can be evaluated by the saddle point method with respect to set of macroscopic parameters $Q_x=[Q_x^{(ab)}], Q_z=[Q_z^{(ab)}]\in\mathbb{R}^{(n+1)\times (n+1)}$:
\begin{align}
    \phi_n &= \frac{1}{\beta}\mathop{\rm extr}_{Q_x, Q_z}[\tilde{g}_{\rm F}(Q_x,Q_z) + \tilde{g}_{\rm G}(Q_x,Q_z) - \tilde{g}_{\rm S}(Q_x,Q_z)], 
    \label{eq:phi_n saddle point representation} \\
    \tilde{g}_{\rm F} &= \lim_{N\to\infty}\frac{1}{N}\log\int 
                            \prod_{a=1}^n 
                            p_{\bm{y}|\bm{z}}(\bm{y}|\bm{z}_a)^\beta
                            p_{\bm{x}_0}(\bm{x}_a)^\beta 
                            q_{\bm{y}|\bm{z}}(\bm{y}|\bm{z}_0)q_{\bm{x}_0}(\bm{x}_0)
    \nonumber \\
    &\times 
                            \prod_{0\le a \le b \le n}\delta(NQ_x^{(ab)}-\bm{x}_a^\top\bm{x}_b)\delta(MQ_z^{(ab)}-\bm{z}_a^\top\bm{z}_b)
                        d^n\bm{x}d^n\bm{z}d\bm{x}_0d\bm{z}_0d\bm{y},
    \nonumber 
\end{align}
where $\tilde{g}_{\rm F}$ can be evaluated analogously as $\tilde{g}_{\rm G}$ and $\tilde{g}_{\rm S}$:
\begin{align}
    \tilde{g}_{\rm F}(Q_x,Q_z) &= \mathop{\rm extr}_{\tilde{Q}_x, \tilde{Q}_z}
    \left[
        \frac{1}{2}{\rm Tr}(Q_x\tilde{Q}_x) + \frac{\delta}{2}{\rm Tr}(Q_z \tilde{Q}_z)
    \right.
    \nonumber \\
    &\left.
        + \log\int e^{-\frac{1}{2}\bm{x}^\top \hat{Q}_x \bm{x}}q_{x_0}(x_0)\prod_{a=1}^n p_{x_0}(x_a)^\beta  dx_0dx_1\dots dx_n
    \right.
    \nonumber \\
    &\left.
        + \delta \log\int e^{-\frac{1}{2}\bm{z}^\top \hat{Q}_z \bm{z}}q_{y|z}(y| z_0)
        \prod_{a=1}^n p_{y|z}(y| z_a)^\beta dz_0dz_1\dots dz_n dy
    \right], 
    \nonumber 
\end{align}
where $\tilde{Q}_x, \tilde{Q}_z\in\mathbb{R}^{(n+1)\times (n+1)}$ are real symmetric matrices, and $\bm{x}=(x_0,x_1,\dots ,x_n), \bm{z}=(z_0, z_1, \dots ,z_n)\in\mathbb{R}^{n+1}$. In general, the extremum conditions for $Q_x, Q_z, \Lambda_x,\Lambda_z, \tilde{Q}_x,$ and $\tilde{Q}_z$ can be written as follows: 
\begin{align*}
    &\left\{
        \begin{array}{c}
             \tilde{Q}_x - Q_x^{-1} + \Lambda_x = 0   
             \vspace{5pt}\\
             \tilde{Q}_z - Q_z^{-1} + \Lambda_z= 0 
        \end{array}
    \right.,
    \quad \mbox{(for $Q_x$ and $Q_z$)}, 
    % \label{eq:saddle_Q_general}
    \\
    &\left\{
        \begin{array}{c}
             Q_x = \mathbb{E}_\lambda\left[\left(\Lambda_x + \lambda \Lambda_z\right)^{-1}\right]   
             \vspace{5pt}\\
             Q_z = \frac{1}{\delta}\mathbb{E}_\lambda\left[\lambda \left(\Lambda_x + \lambda \Lambda_z\right)^{-1}\right]
        \end{array}
    \right.,
    \quad \mbox{(for $\Lambda_x$ and $\Lambda_z$)}, 
    % \label{eq:saddle_Lambda_general}
    \\
    &\left\{
        \begin{array}{c}
            Q_x = \frac{
                \int\bm{x}\bm{x}^\top  e^{-\frac{1}{2}\bm{x}^\top \tilde{Q}_x \bm{x}}q_{x_0}(x_0)\prod_{a=1}^n p_{x_0}(x_a)^\beta dx_0dx_1\dots dx_n
            }{
                \int e^{-\frac{1}{2}\bm{x}^\top \tilde{Q}_x \bm{x}}q_{x_0}(x_0)\prod_{a=1}^n p_{x_0}(x_a)^\beta dx_0dx_1\dots dx_n
            }   
            \vspace{5pt}\\
            Q_z = \frac{
                    \int \bm{z}\bm{z}^\top e^{-\frac{1}{2}\bm{z}^\top \tilde{Q}_z \bm{z}}q_{y|z}(y| z_0)\prod_{a=1}^n p_{y|z}(y| z_a)^\beta dz_0dz_1\dots dz_n dy
                }{
                    \int e^{-\frac{1}{2}\bm{z}^\top \tilde{Q}_z \bm{z}}q_{y|z}(y| z_0)\prod_{a=1}^n p_{y|z}(y| z_a)^\beta dz_0dz_1\dots dz_n dy
                }
        \end{array}
    \right.,
    \quad \mbox{(for $\tilde{Q}_x$ and $\tilde{Q}_z$)}.
    % \label{eq:saddle_Q_tilde_general}
\end{align*}

The key issue is to identify the correct saddle point in (\ref{eq:phi_n saddle point representation}).
Based on the observation that $\tilde{g}_{\rm F}(Q_x, Q_z) + \tilde{g}_{\rm G}(Q_x, Q_z)-\tilde{g}_{\rm S}(Q_x, Q_z)$ in the equation (\ref{eq:phi_n saddle point representation}) is invariant under the permutation of the first to $(n+1)$th lows (and columns) of $Q_x$ and $Q_z$, the RS calculation restricts the candidate of the saddle point to that of the RS form:
\begin{align}
    Q_x &= \left[
            \begin{array}{c|ccc}
                 T_x & m_x & \cdots & m_x  \\
                 \hline
                 m_x & q_x + \frac{\chi_x}{\beta} & & q_x \\
                 \vdots & & \ddots & \\
                 m_x & q_x & & q_x + \frac{\chi_x}{\beta}
            \end{array}
        \right],
        \label{eq:Q_x_RS}
    \\
    Q_z &= \left[
            \begin{array}{c|ccc}
                 T_z & m_z & \cdots & m_z  \\
                 \hline
                 m_z & q_z + \frac{\chi_z}{\beta} & & q_z \\
                 \vdots & & \ddots & \\
                 m_z & q_z & & q_z + \frac{\chi_z}{\beta}
            \end{array}
        \right], 
        \label{eq:Q_z_RS}
    \\
    \tilde{Q}_x &= 
    \left[
        \begin{array}{c|ccc}
             \hat{T}_{1x}& -\beta\hat{m}_{1x}& \cdots & -\beta\hat{m}_{1x} \\
             \hline
             -\beta\hat{m}_{1x} & \beta\hat{Q}_{1x} - \beta^2\hat{\chi}_{1x} & & -\beta^2\hat{\chi}_{1x}  \\
             \vdots & & \ddots &  \\
             -\beta \hat{m}_{1x} & -\beta^2\hat{\chi}_{1x} & &  \beta\hat{Q}_{1x} -\beta^2\hat{\chi}_{1x}
        \end{array}
    \right],
    \label{eq:Q_tilde_x_RS}
    \\
    \tilde{Q}_z &= 
    \left[
        \begin{array}{c|ccc}
             \hat{T}_{2z}& -\beta\hat{m}_{1z}& \cdots & -\beta\hat{m}_{1z} \\
             \hline
             -\beta\hat{m}_{1z} & \beta\hat{Q}_{1z} - \beta^2\hat{\chi}_{1z} & & -\beta^2\hat{\chi}_{1z}  \\
             \vdots & & \ddots &  \\
             -\beta\hat{m}_{1z} & -\beta^2\hat{\chi}_{1z} & &  \beta\hat{Q}_{1z} -\beta^2\hat{\chi}_{1z}
        \end{array}
    \right],
    \label{eq:Q_tilde_z_RS}
    \\
    \Lambda_x &= 
    \left[
        \begin{array}{c|ccc}
             \hat{T}_{2x}& -\beta \hat{m}_{2x}& \cdots & -\beta\hat{m}_{2x} \\
             \hline
             -\beta\hat{m}_{2x} & \beta\hat{Q}_{2x} - \beta^2\hat{\chi}_{2x} & & -\beta^2\hat{\chi}_{2x}  \\
             \vdots & & \ddots &  \\
             -\beta\hat{m}_{2x} & -\beta^2\hat{\chi}_{2x} & &  \beta\hat{Q}_{2x} - \beta^2\hat{\chi}_{2x}
        \end{array}
    \right],
    \label{eq:Lambda_x_RS}
    \\
    \Lambda_z &= 
    \left[
        \begin{array}{c|ccc}
             \hat{T}_{2z}& -\beta\hat{m}_{2z}& \cdots & -\beta\hat{m}_{2z} \\
             \hline
             -\beta\hat{m}_{2z} & \beta\hat{Q}_{2z} - \beta^2\hat{\chi}_{2z}& & -\beta^2\hat{\chi}_{2z}  \\
             \vdots & & \ddots &  \\
             -\beta\hat{m}_{2z} & -\beta^2\hat{\chi}_{2z} & &  \beta\hat{Q}_{2z} - \beta^2\hat{\chi}_{2z}
        \end{array}
    \right],
    \label{eq:Lambda_z_RS}
\end{align}
which is the simplest choice of the saddle point.
This yields the following expression of $\phi_n$:
\begin{align}
    \phi_n &= \mathop{\rm extr}_{T_x, T_z, m_x, m_z, \chi_x, \chi_z, q_x, q_z}\left[\tilde{g}_{\rm F}^{\rm (RS)} + \tilde{g}_{\rm G}^{\rm (RS)} - \tilde{g}_{\rm S}^{\rm (RS)}\right], 
    \label{eq:phi_n_rs_general} \\
    \tilde{g}_{\rm F}^{\rm (RS)} &= 
        \mathop{\rm extr}_{\substack{\hat{T}_{1x}, \hat{T}_{1z}, \hat{m}_{1x}, \hat{m}_{1z}, \\\hat{Q}_{1x}, \hat{Q}_{1z}, \hat{\chi}_{1x}, \hat{\chi}_{1z}}}
        \left[
           \frac{1}{2}T_x \hat{T}_{1x} 
           + \log \int q_{x_0}(x_0)e^{-\frac{\hat{T}_{1x}}{2}x_0^2}dx_0
           +\frac{\delta}{2}T_z \hat{T}_{1z} 
           + \delta \log \int q_{y|z}(y|z_0)e^{-\frac{\hat{T}_{1z}}{2}z_0^2}dx_0dy
        \right.
        \nonumber \\
        &\left.
           + n
            \left(
                - m_x \hat{m}_{1x}  
                +\frac{1}{2}(q_x + \frac{\chi_x}{\beta}) \hat{Q}_{1x} - \frac{1}{2}\chi_x \hat{\chi}_{1x}
            \right)
        % \right.
        % \nonumber \\
        % &\left.
            + n\delta
            \left(
                - m_z \hat{m}_{1z} 
                + \frac{1}{2}(q_z + \frac{\chi_z}{\beta}) \hat{Q}_{1z} - \frac{1}{2}\chi_z \hat{\chi}_{1z}
            \right)
        \right.
        \nonumber \\
        &\left.
            +n\frac{\int q_{x_0}(x_0)e^{-\frac{\hat{T}_{1x}}{2}x_0^2}\phi_x dx_0D\xi_x}{\int q_{x_0}(x_0)e^{-\frac{\hat{T}_{1x}}{2}x_0^2}dx_0}
            +
            n\delta \frac{\int e^{-\frac{\hat{T}_{1z}}{2}z_0^2}q_{y|z}(y| z_0) \phi_z dydz_0 D\xi_z}{\int q_{y|z}(y|z_0)e^{-\frac{\hat{T}_{1z}}{2}z_0^2}dx_0dy}
        \right]+ \mathcal{O}(n^2),
    \label{eq:g_F_rs_general} \\
    % Gauss
    \tilde{g}_{\rm G}^{\rm (RS)} &= \mathop{\rm extr}_{\substack{\hat{T}_{2x}, \hat{T}_{2z}, \hat{m}_{2x}, \hat{m}_{2z}, \\ \hat{Q}_{2x}, \hat{Q}_{2z}, \hat{\chi}_{2x}, \hat{\chi}_{2z}}}
    \left[
        \frac{1}{2}T_x \hat{T}_{2x} +\frac{\delta}{2}T_z \hat{T}_{2z}
        -\frac{1}{2} \log \mathbb{E}_\lambda\left[
            (\hat{T}_{2x} + \lambda\hat{T}_{2z})
        \right]
    \right.
    \nonumber \\
    &\left.
        + n
        \left(
            \frac{1}{2}(q_x + \frac{\chi_x}{\beta}) \hat{Q}_{2x} - \frac{1}{2}\chi_x \hat{\chi}_{2x}- m_x \hat{m}_{2x}
        \right)
    + n\delta
        \left(
            \frac{1}{2}(q_z + \frac{\chi_z}{\beta}) \hat{Q}_{2z} - \frac{1}{2}\chi_z \hat{\chi}_{2z}
            - m_z \hat{m}_{2z}
        \right)
    \right.
    \nonumber \\
    &\left.
        -\frac{n}{2}
        \left\{
            \mathbb{E}_\lambda
            \left[
                \log (\hat{Q}_{2x}+\lambda\hat{Q}_{2z})
            \right]
            -
            \mathbb{E}_\lambda
            \left[
                \frac{
                    \hat{\chi}_{2x} + \lambda \hat{\chi}_{2z}
                }{
                    \hat{Q}_{2x} + \lambda \hat{Q}_{2z}
                }
            \right]
            -
            \mathbb{E}_\lambda
            \left[
                \frac{
                    (\hat{m}_{2x} + \lambda \hat{m}_{2z})^2
                }{
                    (\hat{T}_{2x} + \lambda \hat{T}_{2z})(\hat{Q}_{2x} + \lambda \hat{Q}_{2z})
                }
            \right]
        \right\}
    \right]+ \mathcal{O}(n^2), 
    \label{eq:g_G_rs_general} \\
    % separator
    \tilde{g}_{\rm S}^{\rm (RS)} &= 
        \frac{1}{2}\log  T_x+ \frac{\delta}{2} \log  T_z  +  \frac{n}{2}\left(\frac{\log \chi_x}{\beta} + \frac{q_x}{\chi_x} - \frac{m_x^2}{T_x \chi_x}\right)
        + \frac{n\delta}{2}\left(\frac{\log \chi_z}{\beta} + \frac{q_z}{\chi_z} - \frac{m_z^2}{T_z \chi_z}\right)+ \mathcal{O}(n^2).
    \label{eq:g_S_rs_general}
\end{align}
The condition $\lim_{n\to0}\phi_n=0$ determines $T_x, T_z, \hat{T}_{1x},\hat{T}_{2x}, \hat{T}_{2x}$ and $\hat{T}_{2z}$:
\begin{align}
    T_x &= \int x_0^2 q_{x_0}(x_0) dx_0 ,
    \label{eq:zeroth_order_T_x}
    \\
    T_z &= \frac{1}{\delta}\mathbb{E}_\lambda[\lambda] \int x_0^2 q_{x_0}(x_0) dx_0, 
    \label{eq:zeroth_order_T_z}
    \\
    \hat{T}_{1x} &= 0, 
    \label{eq:zeroth_order_hat_T_1x}
    \\
    \hat{T}_{1z} &= \frac{\delta}{\mathbb{E}_\lambda[\lambda] \int x_0^2 q_{x_0}(x_0) dx_0}, 
    \label{eq:zeroth_order_hat_T_1z}
    \\
    \hat{T}_{2x} &= \frac{1}{T_x},
    \label{eq:zeroth_order_hat_T_2x}
    \\
    \hat{T}_{2z} &= 0.
    \label{eq:zeroth_order_hat_T_2z}
\end{align}
Inserting these conditions into (\ref{eq:phi_n_rs_general})-(\ref{eq:g_S_rs_general}), we obtain  the RS solution which can be extrapolated as $n\to0$:
\begin{align}
    \phi_n = n\times \mathop{\rm extr}_{\substack{m_x,\chi_x,q_x\\m_z,\chi_z,q_z}}\left[g_{\rm F} + g_{\rm G} - g_{\rm S}\right] + \mathcal{O}(n^2)
    \label{eq:phi_n rs}
\end{align}
Inserting the expression (\ref{eq:phi_n rs}) into the replica identity (\ref{eq:replica identity}) yields the RS free energy. 

%%%%%   
\section{de Almeida-Thouless instability condition}
Although the RS form of the saddle point (\ref{eq:Q_x_RS})-(\ref{eq:Lambda_z_RS}) is a natural choice, this choice may lead to wrong free energy \cite{mezard1987spin}. Thus we should investigate the stability of the RS saddle point against RSB. Here, we focus on a local instability scenario, which is termed de Almeida-Thouless instability\cite{dealmeida1978stability}. 

The 1RSB calculation divides the $n$ replicas, which are indexed by $a,b=1,2,\dots ,n$, into $n/\tilde{l}$ groups of identical size $\tilde{l}$ \cite{mezard1987spin} and seeks a saddle point of the following 1RSB form:
\begin{align}
    Q_x^{(ab)} &= 
                \begin{cases}
                    T_x, & a=b=0, \\
                    m_x, & \mbox{\rm $a=0$ or $b=0$}, \\
                    q_x, & \mbox{\rm $a$ and $b$ belong to different groups  ($1\le a,b, a\neq b$)}\\
                    q_x + \Delta_x, & \mbox{$a$ and $b$ belong to an identical group ($1\le a,b, a\neq b$)}   \\
                    q_x + \Delta_x + \frac{\chi_x}{\beta} & \mbox{\rm $1\le a,b$ and $a=b$}
                \end{cases},
                \label{eq:Q_x 1rsb}
    \\
    Q_z^{(ab)} &= 
                \begin{cases}
                    T_z, & a=b=0, \\
                    m_z, & \mbox{\rm $a=0$ or $b=0$}, \\
                    q_z, & \mbox{\rm $a$ and $b$ belong to different groups  ($1\le a,b, a\neq b$)}\\
                    q_z + \Delta_z, & \mbox{$a$ and $b$ belong to an identical group ($1\le a,b, a\neq b$)}   \\
                    q_z + \Delta_z + \frac{\chi_z}{\beta} & \mbox{\rm $1\le a,b$ and $a=b$}
                \end{cases},
    \\
    \tilde{Q}_x^{(ab)} &= 
            \begin{cases}
                \hat{T}_{1x}, & a=b=0, \\
                -\beta \hat{m}_{1x}, & \mbox{\rm $a=0$ or $b=0$}, \\
                -\beta^2 \hat{\chi}_{1x}, & \mbox{\rm $a$ and $b$ belong to different groups  ($1\le a,b, a\neq b$)}\\
                -\beta^2 \hat{\chi}_{1x} - \beta^2 \hat{\Delta}_{1x}, & \mbox{$a$ and $b$ belong to an identical group ($1\le a,b, a\neq b$)}   \\
                \beta \hat{Q}_{1x} - \beta^2 \hat{\chi}_{1x} - \beta^2 \hat{\Delta}_{1x} & \mbox{\rm $1\le a,b$ and $a=b$}
            \end{cases},
    \\
    \tilde{Q}_z^{(ab)} &= 
            \begin{cases}
                \hat{T}_{1z}, & a=b=0, \\
                -\beta \hat{m}_{1z}, & \mbox{\rm $a=0$ or $b=0$}, \\
                -\beta^2 \hat{\chi}_{1z}, & \mbox{\rm $a$ and $b$ belong to different groups  ($1\le a,b, a\neq b$)}\\
                -\beta^2 \hat{\chi}_{1z} - \beta^2 \hat{\Delta}_{1z}, & \mbox{$a$ and $b$ belong to an identical group ($1\le a,b, a\neq b$)}   \\
                \beta \hat{Q}_{1z} - \beta^2 \hat{\chi}_{1z} - \beta^2 \hat{\Delta}_{1z} & \mbox{\rm $1\le a,b$ and $a=b$}
            \end{cases},
    \\
    \Lambda_x^{(ab)} &= 
            \begin{cases}
                \hat{T}_{2x}, & a=b=0, \\
                -\beta \hat{m}_{2x}, & \mbox{\rm $a=0$ or $b=0$}, \\
                -\beta^2 \hat{\chi}_{2x}, & \mbox{\rm $a$ and $b$ belong to different groups  ($1\le a,b, a\neq b$)}\\
                -\beta^2 \hat{\chi}_{2x} - \beta^2 \hat{\Delta}_{2x}, & \mbox{$a$ and $b$ belong to an identical group ($1\le a,b, a\neq b$)}   \\
                \beta \hat{Q}_{2x} - \beta^2 \hat{\chi}_{2x} - \beta^2 \hat{\Delta}_{2x} & \mbox{\rm $1\le a,b$ and $a=b$}
            \end{cases},
    \\
    \Lambda_z^{(ab)} &= 
            \begin{cases}
                \hat{T}_{2z}, & a=b=0, \\
                -\beta \hat{m}_{2z}, & \mbox{\rm $a=0$ or $b=0$}, \\
                -\beta^2 \hat{\chi}_{2z}, & \mbox{\rm $a$ and $b$ belong to different groups  ($1\le a,b, a\neq b$)}\\
                -\beta^2 \hat{\chi}_{2z} - \beta^2 \hat{\Delta}_{2z}, & \mbox{$a$ and $b$ belong to an identical group ($1\le a,b, a\neq b$)}   \\
                \beta \hat{Q}_{2z} - \beta^2 \hat{\chi}_{2z} - \beta^2 \hat{\Delta}_{2z} & \mbox{\rm $1\le a,b$ and $a=b$}
            \end{cases}.
            \label{eq:Lambda_z 1rsb}
\end{align}
In practice, one can label the replicas in such a way that the groups are formed by successive indices $\{1,2,\dots,\tilde{l}\}, \{\tilde{l}+1,\dots,2\tilde{l}\}, \dots, \{n-\tilde{l}+1,\dots, n\}$.
This form of saddle points, in conjunction with the re-scaling of the breaking parameter $\tilde{l}=l/\beta$, yields the following expression of $\phi_n$:
\begin{align}
    \phi_n &= \mathop{\rm extr}_{
        \substack{
            T_x, m_x, q_x, \Delta_x, \chi_x, \\
            T_z, m_z, q_z, \Delta_z, \chi_z
        }
    }\left[\tilde{g}_{\rm F}^{\rm (1RSB)} + \tilde{g}_{\rm G}^{\rm (1RSB)} - \tilde{g}_{\rm S}^{\rm (1RSB)}\right], 
    \label{eq:phi_n_1rsb_general} 
    \\
    % Factorized -------------------------
    \tilde{g}_{\rm F}^{\rm (1RSB)} &= 
        \mathop{\rm extr}_{\substack{
              \hat{T}_{1x}, \hat{m}_{1x}, \hat{\chi}_{1x}, \hat{\Delta}_{1x}, \hat{Q}_{1x}, \\
              \hat{T}_{1z}, \hat{m}_{1z}, \hat{\chi}_{1z}, \hat{\Delta}_{1z}, \hat{Q}_{1z}
            }
        }
        \left[
            % zero-th order
            \frac{1}{2}T_x\hat{T}_{1x}
            + \log \int q_{x_0}(x_0^2)e^{-\frac{\hat{T}_{1x}}{2}x_0^2}dx_0
            +\frac{\delta}{2}T_z \hat{T}_{1z}
            + \delta \log \int q_{y|z}(y|z_0)e^{-\frac{\hat{T}_{1z}}{2}z_0^2}dx_0dy
        \right.
        \nonumber \\
        &\left.
            % x trace
             + n
            \left(
                -m_x\hat{m}_{1x}
                +\frac{1}{2}(q_x + \Delta_x + \frac{\chi_x}{\beta})\hat{Q}_{1x}
                - \frac{l}{2}((q_x+\Delta_x)(\hat{\chi}_{1x} + \hat{\Delta}_{1x}) - q_x \hat{\chi}_{1x})
                -\frac{1}{2}\chi_x (\hat{\chi}_{1x} + \hat{\Delta}_{1x})
            \right)
        \right.
        % z trace
        \nonumber \\ 
        &\left.
            + n\delta
            \left(
                -m_z \hat{m}_{1z} 
                + \frac{1}{2}(q_z + \Delta_z + \frac{\chi_z}{\beta})\hat{Q}_{2x}
                -\frac{l}{2}((q_z + \Delta_z)(\hat{\chi}_{1z}+\hat{\Delta}_{1z})- q_z\hat{\chi}_{1z}) 
                -\frac{1}{2}\chi_z(\hat{\chi}_{1z} + \hat{\Delta}_{1z})
            \right) 
        \right.
        \nonumber \\
        % single body part
        &\left.
            + \frac{n}{l}\frac{
                \int 
                    \left[\log \int e^{l\phi_x^{\rm (1RSB)}(\hat{Q}_{1x}, \hat{\chi}_{1x}, \hat{\Delta}_{1x}, x_0, \xi_x, l)}D\eta_x\right]
                q_{x_0}(x_0)e^{-\frac{\hat{T}_{1x}}{2}x_0^2}dx_0D\xi_x
            }{
                \int q_{x_0}(x_0)e^{-\frac{\hat{T}_{1x}}{2}x_0^2}dx_0
            }
        \right.
        \nonumber \\
        &\left.
            + \frac{n\delta}{l}\frac{
                \int 
                    \left[\log \int e^{l\phi_{z}^{\rm (1RSB)}(\hat{Q}_{1z}, \hat{\chi}_{1z}, \hat{\Delta}_{1z}, z_0, \xi_z, y, l)}D\eta_z \right]
                q_{y|z}(y|z_0)e^{-\frac{\hat{T}_{1z}}{2}z_0^2}dz_0D\xi_zdy
            }{
                \int q_{y|z}(y|z_0)e^{-\frac{\hat{T}_{1z}}{2}z_0^2}dz_0dy
            }
        \right]+ \mathcal{O}(n^2),
    \label{eq:g_F_1rsb_general} 
    \\
    % Gaussian -------------------------
    \tilde{g}_{G}^{\rm (1RSB)} &= 
    \mathop{\rm extr}_{\substack{
              \hat{T}_{2x}, \hat{m}_{2x}, \hat{\chi}_{2x}, \hat{\Delta}_{2x}, \hat{Q}_{2x}, \\
              \hat{T}_{2z}, \hat{m}_{2z}, \hat{\chi}_{2z}, \hat{\Delta}_{2z}, \hat{Q}_{2z}
            }
        }
        \left[
            % zero-th order term
            \frac{1}{2}T_x\hat{T}_{2x}
            +\frac{\delta}{2}T_z \hat{T}_{2z}
            -\frac{1}{2}\mathbb{E}_{\lambda}\left[
                \log (\hat{T}_{2x} + \lambda \hat{T}_{2z})
            \right]
        \right.
        \nonumber \\
        &\left.
            % x trace
            -nm_x\hat{m}_{2x}
                +\frac{n}{2}(q_x + \Delta_x + \frac{\chi_x}{\beta})\hat{Q}_{2x}
            - \frac{nl}{2}((q_x+\Delta_x)(\hat{\chi}_{2x} + \hat{\Delta}_{2x}) - q_x \hat{\chi}_{2x})
            -\frac{n}{2}\chi_x (\hat{\chi}_{2x} + \hat{\Delta}_{2x})
        \right.
        \nonumber \\
        % z trace
        &\left.
            -n\delta m_z \hat{m}_{2z} 
            + \frac{n\delta}{2}(q_z + \Delta_z + \frac{\chi_z}{\beta})\hat{Q}_{2x}
            -\frac{n\delta l}{2}((q_z + \Delta_z)(\hat{\chi}_{2z}+\hat{\Delta}_{2z})- q_z\hat{\chi}_{2z}) 
            -\frac{n\delta}{2}\chi_z(\hat{\chi}_{2z} + \hat{\Delta}_{2z})
        \right.
        \nonumber \\
        % determinant
        &\left.
            -\frac{n}{2}\left(\frac{1}{\beta} - \frac{1}{l}\right)\mathbb{E}_\lambda\left[
                \log \left(
                    \hat{Q}_{2x} + \lambda\hat{Q}_{2z}
                    \right)
            \right]
            -\frac{n}{2l}\mathbb{E}_\lambda\left[
                \log \left(
                    \hat{Q}_{2x} + \lambda\hat{Q}_{2z} -l (\hat{\Delta}_{2x} + \lambda\hat{\Delta}_{2z})
                    \right)
            \right]
        \right.
        \nonumber \\
        &\left.
            +\frac{n}{2}\mathbb{E}_\lambda\left[
                \frac{
                    \hat{\chi}_{2x} + \lambda\hat{\chi}_{2z}
                }{
                    \hat{Q}_{2x} + \lambda\hat{Q}_{2z} -l (\hat{\Delta}_{2x} + \lambda\hat{\Delta}_{2z})
                }
            \right]
            +\frac{n}{2}\mathbb{E}_{\lambda}\left[
                \frac{
                    \left(\hat{m}_{2x} + \lambda\hat{m}_{2z}\right)^2
                }{
                    \left(\hat{T}_{2x} + \lambda\hat{T}_{2z}\right)\left(
                        \hat{Q}_{2x} + \lambda\hat{Q}_{2z} -l (\hat{\Delta}_{2x} + \lambda\hat{\Delta}_{2z})
                    \right)
                }
            \right]
        \right] 
        \nonumber \\
        &+ \mathcal{O}(n^2),
        \label{eq:g_G_1rsb_general} 
        % separator -------------------------
        \\
        \tilde{g}_{\rm S}^{\rm (1RSB)} &= \frac{1}{2}\log  T_x+ \frac{\delta}{2} \log  T_z  
        + \frac{n}{2}\left(\frac{1}{\beta} - \frac{1}{l}\right)\log \chi_x + \frac{n}{2l}\log (\chi_x + l\Delta_x)
        + \frac{n}{2}\frac{q_x}{\chi_x + l\Delta_x} - \frac{n}{2}\frac{m_x^2}{T_x(\chi_x + l\Delta_x)}
        \nonumber \\
        & + \frac{n\delta}{2}\left(\frac{1}{\beta} - \frac{1}{l}\right)\log \chi_z + \frac{n\delta}{2l}\log (\chi_z + l\Delta_z)
        + \frac{n\delta}{2}\frac{q_x}{\chi_z + l\Delta_z} - \frac{n\delta}{2}\frac{m_z^2}{T_z(\chi_z + l\Delta_z)}+ \mathcal{O}(n^2),
        \label{eq:g_S_1rsb_general} 
\end{align}
where 
\begin{align}
    \phi_x^{\rm (1RSB)}(\hat{Q}_{1x}, \hat{\chi}_{1x}, \hat{\Delta}_{1x}, x_0, \xi_x, l) &= \frac{1}{\beta}\log \int e^{-\frac{\beta\hat{Q}_{1x}}{2}x^2 +\beta(\hat{m}_{1x} + \sqrt{\hat{\chi}_{1x}}\xi_{x} + \sqrt{\hat{\Delta}_{1x}}\eta_x)} p_{x_0}(x)^\beta dx
    \label{eq:phi_x_1rsb}
    \\
    \phi_z^{\rm (1RSB)}(\hat{Q}_{1z}, \hat{\chi}_{1z}, \hat{\Delta}_{1z}, z_0, \xi_z, y, l) &=  \frac{1}{\beta}\log \int e^{-\frac{\beta\hat{Q}_{1z}}{2}z^2 +\beta(\hat{m}_{1z} + \sqrt{\hat{\chi}_{1z}}\xi_{z} + \sqrt{\hat{\Delta}_{1z}}\eta_z)} p_{y|z}(z)^\beta dz
    \label{eq:phi_z_1rsb}
\end{align}
Because the zero-th order terms of $n$ in (\ref{eq:g_F_1rsb_general})-(\ref{eq:g_S_1rsb_general}) are identical to the RS calculation (\ref{eq:phi_n_rs_general})-(\ref{eq:g_S_rs_general}), $T_x, T_z, \hat{T}_{1x},\hat{T}_{2x}, \hat{T}_{2x}$ and $\hat{T}_{2z}$ are determined by the equations (\ref{eq:zeroth_order_T_x})-(\ref{eq:zeroth_order_hat_T_2z}), which corresponds to the condition $\lim_{n\to0}\phi_n$. Inserting the equations (\ref{eq:zeroth_order_T_x})-(\ref{eq:zeroth_order_hat_T_2z}) and (\ref{eq:phi_n_1rsb_general})-(\ref{eq:phi_z_1rsb}) to the replica identity (\ref{eq:replica identity}), 1RSB calculation yields the 1RSB free energy $f_{\rm 1RSB}$ as 
\begin{align}
    f_{\rm 1RSB} &= - \mathop{\rm extr}_{\substack{m_x, q_x, \Delta_x, \chi_x, \\
            m_z, q_z, \Delta_z, \chi_z}}\left[g_{\rm F}^{\rm (1RSB)} + g_{\rm G}^{\rm (1RSB)} - g_{\rm S}^{\rm (1RSB)}\right],
            \\
    % Factorized -------------------------
    g_{\rm F}^{\rm (1RSB)} &= \mathop{\rm extr}_{\substack{
              \hat{m}_{1x}, \hat{\chi}_{1x}, \hat{\Delta}_{1x}, \hat{Q}_{1x}, \\
              \hat{m}_{1z}, \hat{\chi}_{1z}, \hat{\Delta}_{1z}, \hat{Q}_{1z}
            }
        }
        \left[
            % x trace
            -m_x\hat{m}_{1x}
            +\frac{1}{2}(q_x + \Delta_x + \frac{\chi_x}{\beta})\hat{Q}_{1x}
            - \frac{l}{2}((q_x+\Delta_x)(\hat{\chi}_{1x} + \hat{\Delta}_{1x}) - q_x \hat{\chi}_{1x})
        \right.    
        \nonumber \\ 
        &\left.
            -\frac{1}{2}\chi_x (\hat{\chi}_{1x} + \hat{\Delta}_{1x})
        % z trace
            -\delta m_z \hat{m}_{1z} 
            + \frac{\delta}{2}(q_z + \Delta_z + \frac{\chi_z}{\beta})\hat{Q}_{2x}
            -\frac{l\delta}{2}((q_z + \Delta_z)(\hat{\chi}_{1z}+\hat{\Delta}_{1z})- q_z\hat{\chi}_{1z}) 
        \right.
        \nonumber \\
        &\left.
            -\frac{\delta}{2}\chi_z(\hat{\chi}_{1z} + \hat{\Delta}_{1z})
        % single body part
            + \frac{1}{l}
                \int \left[
                        \log \int
                            e^{l\phi_x^{\rm (1RSB)}(\hat{Q}_{1x}, \hat{\chi}_{1x}, \hat{\Delta}_{1x}, x_0, \xi_x, l)} 
                        D\eta_x
                    \right] q_{x_0}(x_0)dx_0D\xi_x
        \right.
        \nonumber \\
        &\left.
            + \frac{\delta}{l}
                \int 
                    \left[
                        \log \int
                            e^{l\phi_z^{\rm (1RSB)}(\hat{Q}_{1z}, \hat{\chi}_{1z}, \hat{\Delta}_{1z}, z_0, \xi_z, y, l)}
                        D\eta_z
                    \right]
                q_{y|z}(y|z_0)\sqrt{\frac{\hat{T}_{1z}}{2\pi}}e^{-\frac{\hat{T}_{1z}}{2}z_0^2}dz_0D\xi_zdy
        \right],
        \label{eq:g_F_1rsb} 
    \\
    % Gaussian -------------------------
    g_{G}^{\rm (1RSB)} &= 
    \mathop{\rm extr}_{\substack{
              \hat{m}_{2x}, \hat{\chi}_{2x}, \hat{\Delta}_{2x}, \hat{Q}_{2x}, \\
              \hat{m}_{2z}, \hat{\chi}_{2z}, \hat{\Delta}_{2z}, \hat{Q}_{2z}
            }
        }
        \left[
            % x trace
            -m_x\hat{m}_{2x}
                +\frac{1}{2}(q_x + \Delta_x + \frac{\chi_x}{\beta})\hat{Q}_{2x}
            - \frac{l}{2}((q_x+\Delta_x)(\hat{\chi}_{2x} + \hat{\Delta}_{2x}) - q_x \hat{\chi}_{2x})
        \right.
        \nonumber \\
        &\left.
            -\frac{1}{2}\chi_x (\hat{\chi}_{2x} + \hat{\Delta}_{2x})
        % z trace
            -\delta m_z \hat{m}_{2z} 
            + \frac{\delta}{2}(q_z + \Delta_z + \frac{\chi_z}{\beta})\hat{Q}_{2x}
            -\frac{\delta l}{2}((q_z + \Delta_z)(\hat{\chi}_{2z}+\hat{\Delta}_{2z})- q_z\hat{\chi}_{2z}) 
        \right.
        \nonumber \\
        &\left.     
            -\frac{\delta}{2}\chi_z(\hat{\chi}_{2z} + \hat{\Delta}_{2z})
        % determinant
            -\frac{1}{2}\left(\frac{1}{\beta} - \frac{1}{l}\right)\mathbb{E}_\lambda\left[
                \log \left(
                    \hat{Q}_{2x} + \lambda\hat{Q}_{2z}
                    \right)
            \right]
            -\frac{1}{2l}\mathbb{E}_\lambda\left[
                \log \left(
                    \hat{Q}_{2x} + \lambda\hat{Q}_{2z} -l (\hat{\Delta}_{2x} + \lambda\hat{\Delta}_{2z})
                    \right)
            \right]
        \right.
        \nonumber \\
        &\left.
            +\frac{1}{2}\mathbb{E}_\lambda\left[
                \frac{
                    \hat{\chi}_{2x} + \lambda\hat{\chi}_{2z}
                }{
                    \hat{Q}_{2x} + \lambda\hat{Q}_{2z} -l (\hat{\Delta}_{2x} + \lambda\hat{\Delta}_{2z})
                }
            \right]
            +\frac{T_x}{2}\mathbb{E}_{\lambda}\left[
                \frac{
                    \left(\hat{m}_{2x} + \lambda\hat{m}_{2z}\right)^2
                }{
                    \hat{Q}_{2x} + \lambda\hat{Q}_{2z} -l (\hat{\Delta}_{2x} + \lambda\hat{\Delta}_{2z})
                }
            \right]
        \right] 
        \label{eq:g_G_1rsb},
        \\
        % separator -------------------------
        g_{\rm S}^{\rm (1RSB)} &= 
        \frac{1}{2}\left(\frac{1}{\beta} - \frac{1}{l}\right)\log \chi_x + \frac{1}{2l}\log (\chi_x + l\Delta_x)
        + \frac{1}{2}\frac{q_x}{\chi_x + l\Delta_x} - \frac{1}{2}\frac{m_x^2}{T_x(\chi_x + l\Delta_x)}
        \nonumber \\
        & + \frac{\delta}{2}\left(\frac{1}{\beta} - \frac{1}{l}\right)\log \chi_z + \frac{\delta}{2l}\log (\chi_z + l\Delta_z)
        + \frac{\delta}{2}\frac{q_x}{\chi_z + l\Delta_z} - \frac{\delta}{2}\frac{m_z^2}{T_z(\chi_z + l\Delta_z)}.
        \label{eq:g_S_1rsb} 
\end{align}
Let us denote by $\langle B(x)\rangle_x$ and $\langle C(z)\rangle_z$ expectations
\begin{align}
    \left\langle B(x)\right\rangle_x &= \frac{
        \int 
            B(x)
            e^{
                -\frac{\beta\hat{Q}_{1x}}{2}x^2
                +\beta(\hat{m}_{1x}x_0 + \sqrt{\hat{\chi}_{1x}}\eta_x + \sqrt{\hat{\Delta}_{1x}}\xi_x)x
            }
            p_{x_0}(x_0)^\beta
        dx
    }{
        \int
            e^{
                -\frac{\beta\hat{Q}_{1x}}{2}x^2
                +\beta(\hat{m}_{1x}x_0 + \sqrt{\hat{\chi}_{1x}}\eta_x + \sqrt{\hat{\Delta}_{1x}}\xi_x)x
            }
            p_{x_0}(x_0)^\beta
        dx
    },\\
    \left\langle C(z)\right\rangle_z &= 
    \frac{
        \int C(z)e^{
                -\frac{\beta\hat{Q}_{1z}}{2}z^2
                +\beta(\hat{m}_{1z}z_0 + \sqrt{\hat{\chi}_{1z}}\eta_z + \sqrt{\hat{\Delta}_{1z}}\xi_z)z
            }
            p_{y|z}(y|z)^\beta
        dz
    }{
        \int e^{
                -\frac{\beta\hat{Q}_{1z}}{2}z^2
                +\beta(\hat{m}_{1z}z_0 + \sqrt{\hat{\chi}_{1z}}\eta_z + \sqrt{\hat{\Delta}_{1z}}\xi_z)z
            }
            p_{y|z}(y|z)^\beta
        dz
    },
\end{align}
for arbitrary functions $B(x)$ and $C(z)$.
Then, the extremum conditions are given as follows:
\begin{align}
    0 &= \hat{m}_{1x} + \hat{m}_{2x} - \frac{m_x}{T_x(\chi_x + l\Delta_x)} ,
    \nonumber \\
    0 &= \hat{m}_{1z} + \hat{m}_{2x} - \frac{m_z}{T_z(\chi_z + l\Delta_z)} 
    \nonumber \\
    0 &= \hat{Q}_{1x} + \hat{Q}_{2x} - \frac{1}{\chi_x}, 
    \nonumber \\
    0 &= \hat{Q}_{1z} + \hat{Q}_{2z} - \frac{1}{\chi_z},
    \nonumber \\
    0 &= \hat{Q}_{1x} - l \hat{\Delta}_{1x} + \hat{Q}_{2x} -l\hat{\Delta}_{2x} - \frac{1}{\chi_x +l\Delta_x},
    \label{eq:message_passing_1rsb_x} \\
    0 &= \hat{Q}_{1z} - l \hat{\Delta}_{1z} + \hat{Q}_{2z} -l\hat{\Delta}_{2z} - \frac{1}{\chi_z +l\Delta_z},
    \label{eq:message_passing_1rsb_z} \\
    0 &= \hat{\chi}_{1x} + \hat{\chi}_{2x} - \frac{q_x}{(\chi_x + l\Delta_x)^2} + \frac{m_x^2}{T_x(\chi_x + l\Delta_x^2)},
    \nonumber \\
    0 &= \hat{\chi}_{1z} + \hat{\chi}_{2z} - \frac{q_z}{(\chi_z + l\Delta_z)^2} + \frac{m_z^2}{T_z(\chi_z + l\Delta_z^2)},
    \nonumber \\
    m_x &= \int 
            x_0
            \frac{
                \int \langle x\rangle_x e^{l\phi_{x}^{\rm (1RSB)}}D\eta_x
            }{
                \int e^{l\phi_{x}^{\rm (1RSB)}}D\eta_x
            }
            dq_{x_0}(x_0)dx_0D\xi_x,
    \nonumber \\
    q_x &= \int 
            \left[
                \frac{
                    \int \langle x\rangle_x e^{l\phi_{x}^{\rm (1RSB)}}D\eta_x
                }{
                    \int e^{l\phi_{x}^{\rm (1RSB)}}D\eta_x
                }
            \right]^2
            dq_{x_0}(x_0)dx_0D\xi_x,
    \nonumber \\
    \frac{\chi_x}{\beta} &= 
        \int 
            \frac{
                \int [\langle x^2\rangle_x - \langle x\rangle_x^2] e^{l\phi_{x}^{\rm (1RSB)}}D\eta_x
            }{
                \int e^{l\phi_{x}^{\rm (1RSB)}}D\eta_x
            }
        dq_{x_0}(x_0)dx_0D\xi_x,
    \nonumber \\
    \Delta_x &= 
        \int
            \left[
                \frac{
                    \int \langle x\rangle_x^2 e^{l\phi_{x}^{\rm (1RSB)}}D\eta_x
                }{
                    \int e^{l\phi_{x}^{\rm (1RSB)}}D\eta_x
                }
                -
                \left(
                    \frac{
                        \int \langle x\rangle_x e^{l\phi_{x}^{\rm (1RSB)}}D\eta_x
                    }{
                        \int e^{l\phi_{x}^{\rm (1RSB)}}D\eta_x
                    }
                \right)^2
            \right]
        dq_{x_0}(x_0)dx_0D\xi_x,
    \label{eq:delta_x_factorized} \\
    m_z &= \int 
            z_0
            \frac{
                \int \langle z\rangle_z e^{l\phi_{z}^{\rm (1RSB)}}D\eta_z
            }{
                \int e^{l\phi_{z}^{\rm (1RSB)}}D\eta_z
            }
            dq_{y|z}(y|z_0)dz_0dyD\xi_z,
    \nonumber \\
    q_z &= \int 
            \left[
                \frac{
                    \int \langle z\rangle_z e^{l\phi_{z}^{\rm (1RSB)}}D\eta_z
                }{
                    \int z^{l\phi_{z}^{\rm (1RSB)}}D\eta_z
                }
            \right]^2
            dq_{y|z}(y|z_0)dz_0dyD\xi_z,
    \nonumber \\
    \frac{\chi_z}{\beta} &= 
            \int 
                \frac{
                    \int [\langle z^2\rangle_z - \langle z\rangle_z^2] e^{l\phi_{z}^{\rm (1RSB)}}D\eta_z
                }{
                    \int z^{l\phi_{z}^{\rm (1RSB)}}D\eta_z
                }
            dq_{y|z}(y|z_0)dz_0dyD\xi_z,
    \nonumber \\
    \Delta_z &= 
            \int 
                \left[
                    \frac{
                        \int \langle z\rangle_z^2 e^{l\phi_{z}^{\rm (1RSB)}}D\eta_z
                    }{
                        \int z^{l\phi_{z}^{\rm (1RSB)}}D\eta_z
                    }
                    -
                    \left(
                        \frac{
                            \int \langle z\rangle_z e^{l\phi_{z}^{\rm (1RSB)}}D\eta_z
                        }{
                            \int z^{l\phi_{z}^{\rm (1RSB)}}D\eta_z
                        }
                    \right)^2
                \right]
            dq_{y|z}(y|z_0)dz_0dyD\xi_z,
    \label{eq:delta_z_factorized} \\
    m_x &= T_x\mathbb{E}_{\lambda}\left[
        \frac{
            \hat{m}_{2x} + \lambda\hat{m}_{2z}
        }{
            \hat{Q}_{2x} + \lambda\hat{Q}_{2z} - l (\hat{\Delta}_{2x} + \lambda \hat{\Delta}_{2z})
        }
    \right],
    \nonumber \\
    q_x &= \mathbb{E}_\lambda\left[
        \frac{
            \hat{\chi}_{2x} + \lambda\hat{\chi}_{2z}
        }{
            (\hat{Q}_{2x} + \lambda\hat{Q}_{2z} - l (\hat{\Delta}_{2x} + \lambda \hat{\Delta}_{2z}))^2
        }
    \right]
    +
    T_x\mathbb{E}_{\lambda}\left[
        \left(
            \frac
            {
                \hat{m}_{2x} + \lambda\hat{m}_{2z}
            }{
                \hat{Q}_{2x} + \lambda\hat{Q}_{2z} - l (\hat{\Delta}_{2x} + \lambda \hat{\Delta}_{2z})
            }
        \right)^2
    \right],
    \nonumber \\
    \chi_x &= \mathbb{E}_{\lambda}\left[
        \frac{1}{\hat{Q}_{2x} + \lambda\hat{Q}_{2z}}
    \right], 
    \nonumber \\
    \Delta_x &= \frac{1}{l}\left(
            \mathbb{E}_{\lambda}\left[
                \frac{1}{\hat{Q}_{2x} + \lambda\hat{Q}_{2z} - l (\hat{\Delta}_{2x} + \lambda \hat{\Delta}_{2z})}
            \right]
            -
            \mathbb{E}_{\lambda}\left[
                \frac{1}{\hat{Q}_{2x} + \lambda \hat{Q}_{2z}}
            \right]
        \right),
    \label{eq:delta_x_gaussian} \\
    m_z &= \frac{1}{\delta}\mathbb{E}_{\lambda}\left[
        \frac{
            \lambda(\hat{m}_{2x} + \lambda\hat{m}_{2z})
        }{
             \hat{Q}_{2x} + \lambda\hat{Q}_{2z} - l (\hat{\Delta}_{2x} + \lambda \hat{\Delta}_{2z})
        }
    \right],
    \nonumber \\
    q_x &= \frac{1}{\delta}\mathbb{E}_\lambda\left[
        \frac{
            \lambda(\hat{\chi}_{2x} + \lambda\hat{\chi}_{2z})
        }{
            (\hat{Q}_{2x} + \lambda\hat{Q}_{2z} - l (\hat{\Delta}_{2x} + \lambda \hat{\Delta}_{2z}))^2
        }
    \right]
    +
    \frac{T_x}{\delta}\mathbb{E}_{\lambda}\left[
        \lambda\left(
            \frac
            {
                \hat{m}_{2x} + \lambda\hat{m}_{2z}
            }{
                \hat{Q}_{2x} + \lambda\hat{Q}_{2z} - l (\hat{\Delta}_{2x} + \lambda \hat{\Delta}_{2z})
            }
        \right)^2
    \right],
    \nonumber \\
    \chi_z &= \frac{1}{\delta}\mathbb{E}_{\lambda}\left[
        \frac{\lambda}{\hat{Q}_{2x} + \lambda\hat{Q}_{2z}}
    \right], 
    \nonumber \\
    \Delta_z &= \frac{1}{\delta l}\left(
        \mathbb{E}_{\lambda}\left[
            \frac{\lambda}{\hat{Q}_{2x} + \lambda\hat{Q}_{2z} - l (\hat{\Delta}_{2x} + \lambda \hat{\Delta}_{2z})}
        \right]
        -
        \mathbb{E}_{\lambda}\left[
            \frac{\lambda}{\hat{Q}_{2x}+\lambda\hat{Q}_{2z}}
        \right]
    \right).
    \label{eq:delta_z_gaussian}
\end{align}
In the limit of $n\to0$, the 1RSB calculation imposes the group size $\tilde{l}$, which was originally introduced as an integer between $1$ and $n$, to be a real number $\tilde{l}\in[0,1]$. For any $l\in[0,\beta]$, the condition $\Delta_x=\Delta_z=\hat{\Delta}_{1x}=\hat{\Delta}_{1z}=\hat{\Delta}_{2x}=\hat{\Delta}_{2z}=0$ reproduces the RS solution. RSB means that these parameters are not zero.
Hence, one can check the validity of the RS solution by examining the stability of the solution of $\Delta_x=\Delta_z=\hat{\Delta}_{1x}=\hat{\Delta}_{1z}=\hat{\Delta}_{2x}=\hat{\Delta}_{2z}=0$ under the 1RSB calculation. Around the RS solution, the extremum condition (\ref{eq:message_passing_1rsb_x})-(\ref{eq:delta_z_gaussian}) are expanded as 
\begin{align}
    \left[
        \begin{array}{c}
             \Delta_x  \\
             \Delta_z
        \end{array}
    \right]
    &\simeq
    \left[
        \begin{array}{cc}
             \chi_x^{(2)} & 0  \\
             0 &  \chi_z^{(2)}
        \end{array}
    \right]
    \left[
        \begin{array}{c}
             \hat{\Delta}_{1x}  \\
             \hat{\Delta}_{1z} 
        \end{array}
    \right],
    \label{eq:delta_linearlized_factorized}
    \\
    \left[
        \begin{array}{c}
             \hat{\Delta}_{1x}  \\
             \hat{\Delta}_{1z}
        \end{array}
    \right]
    &\simeq
        \left[
        \begin{array}{cc}
             \frac{1}{\chi_x^2} & 0  \\
             0 &  \frac{1}{\chi_z^2}
        \end{array}
    \right]
    \left[
        \begin{array}{c}
             \Delta_x  \\
             \Delta_z
        \end{array}
    \right]
    - \left[
        \begin{array}{c}
             \hat{\Delta}_{2x}  \\
             \hat{\Delta}_{2z}
        \end{array}
    \right],
    \label{eq:delta_1_hat_linearlized_gaussian}
    \\
     \left[
        \begin{array}{c}
             \Delta_x  \\
             \Delta_z
        \end{array}
    \right]
    &\simeq
    \left[
        \begin{array}{cc}
             \zeta_0 & \zeta_1  \\
             \frac{\zeta_1}{\delta} &  \frac{\zeta_2}{\delta}
        \end{array}
    \right]
    \left[
        \begin{array}{c}
             \hat{\Delta}_{2x}  \\
             \hat{\Delta}_{2z}
        \end{array}
    \right]
    .
    \label{eq:delta_linearlized_gaussian}
\end{align}
Solving (\ref{eq:delta_linearlized_gaussian}) for $\hat{\Delta}_{2x}$ and $\hat{\Delta}_{2z}$ gives
\begin{align}
    \left[
        \begin{array}{c}
             \hat{\Delta}_{2x}  \\
             \hat{\Delta}_{2z}
        \end{array}
    \right]
    &\simeq
    \left[
        \begin{array}{cc}
             \frac{1}{\chi_x^2} - \frac{\zeta_2}{\zeta_0\zeta_2-\zeta_1^2}
             &
             \frac{\zeta_1}{\zeta_0\zeta_2-\zeta_1^2}
             \\
             \frac{\delta\zeta_1}{\zeta_0\zeta_2-\zeta_1^2}
             &
             \frac{1}{\chi_z^2} - \frac{\delta \zeta_0}{\zeta_0\zeta_2-\zeta_1^2}
        \end{array}
    \right]
    \left[
        \begin{array}{c}
             \Delta_{2x}  \\
             \Delta_{2z}
        \end{array}
    \right]
    =
    \left[
        \begin{array}{cc}
             2\frac{\partial^2 \mathcal{F}}{\partial \chi_x^2} & 2\frac{\partial^2 \mathcal{F}}{\partial\chi_x\partial \chi_z}  \\
             \frac{2}{\delta}\frac{\partial^2 \mathcal{F}}{\partial\chi_x\partial \chi_z} &  \frac{2}{\delta}\frac{\partial^2 \mathcal{F}}{\partial \chi_z^2}
        \end{array}
    \right]
    \left[
        \begin{array}{c}
             \Delta_{2x}  \\
             \Delta_{2z}
        \end{array}
    \right].
    \label{eq:delta_2_hat_linearlized_gaussian}
\end{align}
Inserting (\ref{eq:delta_2_hat_linearlized_gaussian}) and (\ref{eq:delta_1_hat_linearlized_gaussian}) into (\ref{eq:delta_linearlized_factorized}) yields
\begin{align}
    \left[
        \begin{array}{c}
             \Delta_x  \\
             \Delta_z
        \end{array}
    \right]
    \simeq 
    &\left[
        \begin{array}{cc}
            2\frac{\partial^2 \mathcal{F}}{\partial \chi_x^2}
            \chi_x^{(2)} 
            &
            2\frac{\partial^2 \mathcal{F}}{\partial\chi_x\partial\chi_z} 
            \chi_x^{(2)}
            \\
             \frac{2}{\delta}\frac{\partial^2 \mathcal{F}}{\partial \chi_x\partial\chi_z}
            \chi_z^{(2)} 
            &
            \frac{2}{\delta}\frac{\partial^2 \mathcal{F}}{\partial\chi_z^2} 
            \chi_z^{(2)}
        \end{array}
    \right]
    \left[
        \begin{array}{c}
             \Delta_x  \\
             \Delta_z
        \end{array}
    \right],
\end{align}
which indicates that the solution $\Delta_x=\Delta_z=0$ is unstable if the largest eigenvalue of the matrix
\begin{equation*}
    \left[
        \begin{array}{cc}
            2\frac{\partial^2 \mathcal{F}}{\partial \chi_x^2}
            \chi_x^{(2)} 
            &
            2\frac{\partial^2 \mathcal{F}}{\partial\chi_x\partial\chi_z} 
            \chi_x^{(2)}
            \\
            \frac{2}{\delta}\frac{\partial^2 \mathcal{F}}{\partial \chi_x\partial\chi_z}
            \chi_z^{(2)} 
            &
            \frac{2}{\delta}\frac{\partial^2 \mathcal{F}}{\partial\chi_z^2} 
            \chi_z^{(2)}
        \end{array}
    \right],
\end{equation*}
is greater than $1$. This condition yields the AT instability condition (\ref{eq:AT-replica}).

%%%%%
\section{Microscopic instability of VAMP}
 Let us define $\mathcal{O}(\epsilon^\alpha), \alpha>0$ as $\mathcal{O}(\epsilon_x^\alpha) + \mathcal{O}(\epsilon_z^\alpha)$, and denote by $\bm{v}\circ\bm{w}=[v_iw_i]$ as the Hadamard product for vectors $\bm{v}=[v_i],\bm{w}=[w_i]$. We are interested in how the small perturbations $\epsilon_x\bm{\eta}_x$ and $\epsilon_z\bm{\eta}_z$ evolve after single iteration of the Algorithm \ref{algo:VAMP}. To this aim, we expand the each step of the Algorithm \ref{algo:VAMP} at the leading order of $\epsilon$.

The equations in the factorized part (line \ref{line:factorized x} and \ref{line:factorized z} in Algorithm \ref{algo:VAMP}) are expanded as follows:
\begin{align*}
    \hat{\bm{x}}_1^{(t)} &= \hat{\bm{x}}_1 + \sqrt{\epsilon_x}\frac{\partial \hat{\bm{x}}_1}{\partial \bm{h}_{1x}} \circ \bm{\eta}_x + \mathcal{O}(\epsilon), \\
    \hat{\bm{z}}_1^{(t)} &= \hat{\bm{z}}_1 + \sqrt{\epsilon_z}\frac{\partial \hat{\bm{z}}_1}{\partial \bm{h}_{1z}} \circ \bm{\eta}_z + \mathcal{O}(\epsilon), \\
    \chi_{1x}^{(t)} &= \chi_{1x} + \mathcal{O}(\epsilon), \\
    \chi_{1z}^{(t)} &= \chi_{1z} + \mathcal{O}(\epsilon).
\end{align*}
Then, the equations in the message passing part (line \ref{line:message FtoG 1} and \ref{line:message FtoG 2} in Algorithm \ref{algo:VAMP}) are expanded as:
\begin{align}
    \bm{h}_{2x}^{(1)} &= \bm{h}_{2x} + \sqrt{\epsilon_x}\left(\frac{1}{\chi_{1x}}\frac{\partial \hat{\bm{x}}_1}{\partial \bm{h}_{1x}} - \bm{1}_N\right) \circ \bm{\eta}_x + \mathcal{O}(\epsilon), 
    \nonumber \\
    &\equiv \bm{h}_{2x} + \sqrt{\epsilon_x}\bm{\eta}_{{\rm G}, x} + \mathcal{O}(\epsilon)
    \nonumber \\
    \bm{h}_{2z}^{(1)} &= \bm{h}_{2z} + \sqrt{\epsilon_x}\left(\frac{1}{\chi_{1z}}\frac{\partial \hat{\bm{z}}_1}{\partial \bm{h}_{1z}} - \bm{1}_M\right) \circ \bm{\eta}_z + \mathcal{O}(\epsilon), 
    \nonumber \\
    &\equiv \bm{h}_{2z} + \sqrt{\epsilon_z}\bm{\eta}_{{\rm G}, z} + \mathcal{O}(\epsilon)
    \nonumber\\
    \hat{Q}_{2x}^{(1)} &= \hat{Q}_{2x} + \mathcal{O}(\epsilon), 
    \nonumber \\
    \hat{Q}_{2z}^{(1)} &= \hat{Q}_{2z} + \mathcal{O}(\epsilon),
    \nonumber 
\end{align}
where $\bm{1}_N = (1,1,\dots,1)^\top\in\mathbb{R}^N$ and $\bm{1}_M=(1,1,\dots ,1)^\top\in\mathbb{R}^M$.
Similarly, the Gaussian part (line \ref{line:Gaussian x 1}-\ref{line:Gaussian z 2} in Algorithm \ref{algo:VAMP}) can be expanded as:
\begin{align}
    \hat{\bm{x}}_2^{(1)} &= \hat{\bm{x}}_2 + \sqrt{\epsilon_x}K^{-1}\bm{\eta}_{{\rm G},x} + \sqrt{\epsilon_z} K^{-1}A^\top \bm{\eta}_{{\rm G}, z} + \mathcal{O}(\epsilon), 
    \nonumber \\
    \hat{\bm{z}}_2^{(1)} &= \hat{\bm{z}}_2 + \sqrt{\epsilon_x}AK^{-1}\bm{\eta}_{{\rm G},x} + \sqrt{\epsilon_z} AK^{-1}A^\top \bm{\eta}_{{\rm G}, z} + \mathcal{O}(\epsilon), 
    \nonumber \\
    \chi_{2x}^{(1)} &= \chi_{2x} + \mathcal{O}(\epsilon), 
    \nonumber \\
    \chi_{2z}^{(1)} &= \chi_{2z} + \mathcal{O}(\epsilon).
    \nonumber
\end{align}
Finally, the message passing part (line \ref{line:message GtoF 1} and \ref{line:message GtoF 2}) are written as follows:
\begin{align}
    \bm{h}_{1x}^{(2)} &= \bm{h}_{2x} + \sqrt{\epsilon_x}\left(\frac{1}{\chi_{2x}}K^{-1} - I_N\right)\bm{\eta}_{{\rm G}, x}
    % \nonumber \\
    % &
    + \sqrt{\epsilon_z} K^{-1} A^\top \bm{\eta}_{{\rm G},z} + \mathcal{O}(\epsilon)
    \label{eq:h_1x perturbation}\\
    \bm{h}_{1z}^{(2)} &= \bm{h}_{2z} + \sqrt{\epsilon_x}AK^{-1}\bm{\eta}_{{\rm G}, x}
    % \nonumber \\
    % &
    + \sqrt{\epsilon_z} \left(\frac{1}{\chi_{2z}}AK^{-1}A^\top - I_M\right) \bm{\eta}_{{\rm G},z} + \mathcal{O}(\epsilon)
    \label{eq:h_1z perturbation},
    \\
    \hat{Q}_{1x}^{(2)} &= \hat{Q}_{1x} + \mathcal{O}(\epsilon), 
    \nonumber \\
    \hat{Q}_{1z}^{(2)} &= \hat{Q}_{1z} + \mathcal{O}(\epsilon).
    \nonumber 
\end{align}
After single iteration of the Algorithm \ref{algo:VAMP}, the perturbation affect only on $\bm{h}_{1x}^{(2)}$ and $\bm{h}_{1z}^{(2)}$ at the leading order.
Using the independence of $\bm{\eta}_x$ and $\bm{\eta}_z$, the variances of the perturbation terms in (\ref{eq:h_1x perturbation}) and (\ref{eq:h_1z perturbation}) can be written as 
\begin{equation}
     \left(\frac{\zeta_0}{(\chi_{x})^2} - 1\right)\left(\frac{\chi_{x}^{(2)} }{\chi_{x}^2}- 1\right)\epsilon_x + \frac{\zeta_1}{\chi_{x}^2}\left(\frac{\chi_{z}^{(2)}}{\chi_{z}^2} - 1\right)\epsilon_z,
     \nonumber
\end{equation}
and 
\begin{equation}
    \frac{\zeta_1}{\delta\chi_{z}^2}\left(\frac{\chi_{x}^{(2)}}{\chi_{x}^2} - 1\right) \epsilon_x +
    \left(\frac{\zeta_2}{\chi_{z}^2} - 1\right)\left(\frac{\chi_{z}^{(2)}}{\chi_{z}^2} - 1\right)\epsilon_z\nonumber
\end{equation}
Thus, the variances of the perturbation terms grow exponentially by the VAMP iterations if the largest eigenvalue of the matrix
\begin{align}
    \left[
        \begin{array}{cc}
             \left(\frac{\zeta_0}{(\chi_{x})^2} - 1\right)\left(\frac{\chi_{x}^{(2)} }{\chi_{x}^2}- 1\right)
             &
             \frac{\zeta_1}{\chi_{x}^2}\left(\frac{\chi_{z}^{(2)}}{\chi_{z}^2} - 1\right)
             \\
             \frac{\zeta_1}{\delta\chi_{z}^2}\left(\frac{\chi_{x}^{(2)}}{\chi_{x}^2} - 1\right) 
             &
             \left(\frac{\zeta_2}{\chi_{z}^2} - 1\right)\left(\frac{\chi_{z}^{(2)}}{\chi_{z}^2} - 1\right)
        \end{array}
    \right],
    \nonumber 
\end{align}
is greater than $1$. This condition yields the microscopic instability condition (\ref{eq:micro instability}).

\section*{Acknowledgments}

This work was supported by JSPS KAKENHI Grant Numbers 19J10711, 17H00764, and JST CREST Grant Number JPMJCR1912, Japan.

%%%%%%
%% To balance the columns at the last page of the paper use this
%% command:
%%
% \enlargethispage{-1.2cm} 
%%
%% If the balancing should occur in the middle of the references, use
%% the following trigger:
%%
% \IEEEtriggeratref{3}
%%
%% which triggers a \newpage (i.e., new column) just before the given
%% reference number. Note that you need to adapt this if you modify
%% the paper.  The "triggered" command can be changed if desired:
%%
% \IEEEtriggercmd{\enlargethispage{-20cm}}
%%
%%%%%%

%%%%%
% References:
% We recommend the usage of BibTeX:
%
\bibliographystyle{IEEEtran}
\bibliography{references}

%
% where we here have assume the existence of the files
% definitions.bib and bibliofile.bib.
% BibTeX documentation can be obtained at:

\end{document}